\documentclass[12.0pt]{article}
\usepackage[a4paper, total={6.2in,9.6in}]{geometry}
\usepackage{fancyhdr}
\usepackage{graphicx}
\usepackage{amsmath, mathrsfs}
\usepackage{multirow}
\usepackage{slashed}
\usepackage{graphicx,rotate,multicol}
\usepackage{wrapfig}
\usepackage{blindtext}
\usepackage{enumitem}
\usepackage{epsfig}
\usepackage{latexsym, amssymb} 
\usepackage[colorlinks=true,
linkcolor=red,
urlcolor=blue,
citecolor=blue]{hyperref}
\usepackage{cite}
\usepackage{ctable}
\usepackage{footnote}
\usepackage{float}
\usepackage[caption=false]{subfig}
\usepackage{xcolor,cancel}
\newcommand{\beq}{\begin{equation}} 
\newcommand{\eeq}{\end{equation}} 
\newcommand{\ba}{\begin{array}}  
\newcommand{\ea}{\end{array}} 
\newcommand{\bea}{\begin{eqnarray}}  
\newcommand{\eea}{\end{eqnarray} }  
\newcommand{\be}{\begin{eqnarray}}  
\newcommand{\ee}{\end{eqnarray} }  
\newcommand{\bal}{\begin{align}}
\newcommand{\eal}{\end{align}}   
\newcommand{\bi}{\begin{itemize}}  
\newcommand{\ei}{\end{itemize}}  
\newcommand{\ben}{\begin{enumerate}}  
\newcommand{\een}{\end{enumerate}}  
\newcommand{\bc}{\begin{center}}
\newcommand{\ec}{\end{center}} 
\newcommand{\bt}{\begin{table}}
\newcommand{\et}{\end{table}}  
\newcommand{\btb}{\begin{tabular}}
\newcommand{\etb}{\end{tabular}}

\newcommand{\cO}{{\mathcal O}} 

\renewcommand{\thesection}{\Roman{section}}

\newcommand{\nc}{\newcommand}

\nc{\scm}{\scriptscriptstyle\mathrm}
\nc{\f}{\frac}
\nc{\baa}{\begin{array}}      \nc{\eaa}{\end{array}}
\nc{\bit}{\begin{itemize}}    \nc{\eit}{\end{itemize}}
\nc{\bce}{\begin{center}}     \nc{\ece}{\end{center}}
\nc{\bfl}{\begin{flushright}} \nc{\efl}{\end{flushright}}
\nc{\vp}{\varphi}
\nc{\tvp}{\widetilde{\varphi}}
\nc{\D}{\mbox{$\not\!\!D$}}
\nc{\Db}{\mbox{${\raisebox{2mm}{\boldmath ${}^\leftarrow$}\hspace{-4mm} D}$}}
\nc{\Dfb}{\mbox{$\raisebox{2mm}{\boldmath ${}^\leftrightarrow$}\hspace{-4mm} D$}}
\nc{\vpj }{\mbox{${\vp^\dag i\,\raisebox{2mm}{\boldmath ${}^\leftrightarrow$}\hspace{-4mm} D_\mu\,\vp}$}}
\nc{\vpjt}{\mbox{${\vp^\dag i\,\raisebox{2mm}{\boldmath ${}^\leftrightarrow$}\hspace{-4mm} D_\mu^{\,I}\,\vp}$}}

\begin{document}
\vskip 30pt  
 
\begin{center}  
{\Large \bf Following the trail of new physics via VBF Higgs signal at the Large Hadron Collider} \\
\vspace*{1cm}  
\renewcommand{\thefootnote}{\fnsymbol{footnote}}  
{\sf Tisa Biswas$^1$ \footnote{email: tibphy\_rs@caluniv.ac.in}},
{\sf Anindya Datta$^1$ \footnote{email: adphys@caluniv.ac.in}},  
{\sf Biswarup Mukhopadhyaya$^2$ \footnote{email: biswarup@iiserkol.ac.in} }\\  
\vspace{10pt}  
{ {\em $^1$Department of Physics, University of Calcutta,  
92 Acharya Prafulla Chandra Road, \\ Kolkata - 700009, India}
\\
{\em $^2$Department of Physical Sciences, Indian Institute of Science Education and Research,\\ Mohanpur - 741246, India\\
}}
\normalsize  
\end{center} 

\bigskip
\begin{abstract}
 We investigate the modification of the Higgs signals from vector boson fusion at the LHC arising from higher-dimensional effective operators involving quarks, electroweak gauge bosons and the 125-GeV scalar discovered in 2012. Taking a few of the admissible dimension-6 operators as illustration, we work within the framework of the Standard Model Effective Field Theory (SMEFT) and identify kinematic variables that can reflect the presence of such effective operators. The useful variables turn out to be the geometric mean of the transverse momenta of the two forward jets produced in VBF and the rapidity difference between the two forward jets. We identify the shift in event population caused by the effective operators in the same, spanned by the above kinematic variables. Minimum values of the Wilson coefficients of the chosen dimension-6 operators are identified, for which they can be probed at the $3\sigma$ level in the high luminosity run of the LHC at 14 TeV. Projected exclusion limits on some of the couplings, obtained from our analysis, can significantly improve the limits on such couplings derived from  electroweak precision data. 
\end{abstract}

\texttt{Key Words:~~Higgs Boson, Vector Boson Fusion, Large Hadron Collider, Standard Model Effective Field theory, HL-LHC} 
\renewcommand{\thesection}{\Roman{section}}  
\setcounter{footnote}{0}  
\renewcommand{\thefootnote}{\arabic{footnote}}  

\bigskip
\section{Introduction}
\label{sec:Intro}
The Standard Model (SM) of Particle Physics has been proven to be remarkably successful in explaining most observations, starting from  low-energy observables in  weak decays to multi-particle production at the Large Hadron Collider (LHC).  The last knot in the SM thread appears to be tied now, when it is establised that the 125 GeV scalar discovered\cite{Aad:2012tfa,Chatrchyan:2012xdj} in 2012 is responsible (at least dominantly) in electroweak symmetry breaking and mass generation. Unfortunately, the SM is unable to account for the measured relic density, non-zero neutrino mass, baryon - anti-baryon asymmetry etc. along with aesthetic issues like fine tuning in Higgs boson mass. It is a general consensus that it is  a part of a more fundamental and complete theory which will be revealed to us at some higher energy scale. With the hope of testing the limitations of the SM, experimental results from the LHC has not revealed any hint of any such theory so far. However, the LHC is more than a particle discovery machine. We are entering an era of precision measurements as the high luminosity run is close to its take-off.  Uncovering traces of new Physics in the high luminosity run  is still a well-founded hope, if theoretical predictions can capture the deviation from SM with appropriate  parametrisation. It is in this spirit that effective operators are introduced, involving the SM fields. Such operators encapsulate contributions potentially arising from physics lying beyond the reach of direct searches, by modifying kinematical features of various final states especially in the high-energy tails. Therein lies the essential role of Standard Model Effective Field Theory (SMEFT)\cite{Weinberg:1978kz,Buchmuller,Leung:1984ni,Henning:2014wua,Dawson:2018dcd}.

In this formulation, it is assumed that if there is any new Physics associated with the electroweak symmetry breaking sector, the Higgs observed at the LHC is still a part of an $SU(2)_{L}$ doublet, the SM gauge invariance holds and no additional light degrees of freedom relevant to the Higgs observables, are present in the spectrum. SMEFT interactions can be expressed as an operator expansion in inverse powers of a high energy scale, $\Lambda$,
\begin{equation}
\mathcal{L}_{\rm SMEFT}=
\mathcal{L}_{\rm SM}+\sum_{i,n}{C_i^{(n)}\over\Lambda^{n-4}}\mathcal{O}_i^{(n)}+\ldots\,
\label{eq:smeft}
\end{equation}

Here, the leading order term is the complete SM Lagrangian and  $\mathcal{O}_i^{(n)}$ are operators with mass dimension-$n$, constructed from the SM fields and all of the BSM physics effects reside in the coefficient  $C_i^{(n)}$. The leading order new Physics effects that are associated with the EFT operators, apart from dimension-5 operator that contributes to neutrino masses, are of dimension-6. Neglecting flavour, there are 59 possible operators at dimension-6 \cite{Grzadkowski:2010es,Giudice:2007fh}.

At the LHC, the second most copious source for SM Higgs,  next to the gluon fusion channel, is the vector boson fusion (VBF) mechanism\cite{Hankele:2006ma}. This process has a rich kinematic structure with two forward tagging jets and little hadronic activity in the rapidity interval between them, resulting in clean samples of signal events and allowing for measuring the properties of Higgs with gauge bosons and fermions. Modifications from anomalous HVV coupling to VBF have been evaluated in Refs.\cite{Englert:2014uua,Banerjee:2013apa, Ellis:2014dva,Djouadi:2013yb,Biswal:2012mp,Amar:2014fpa,LHCHiggsCrossSectionWorkingGroup:2016ypw,DelDuca:2001fn}. Constraints on the higher dimensional operators have been extensively studied on the basis of electroweak precision test
and global fits of Higgs data in Refs.\cite{Peskin:1991sw,Masso:2012eq,Farina:2016rws,Falkowski:2013dza,Einhorn:2013tja,Ellis:2018gqa,Cepeda:2019klc}.

In this paper, we show that the dynamics of the forward tagging jets is sensitive to  some additional effective interactions leading to Higgs boson production using the same final states as those studied for the VBF channel. We will critically look at the kinematic distributions of the tagging jets instead of looking for the Higgs decay products in the rapidity gap between two forward jets.   
 We have used the  $\gamma \gamma$ decay channel of the Higgs boson in our analysis. However, our main focus will be  on the  jet observables and our method does not crucially depend on the Higgs decay modes. 
Consequently, other decay modes of Higgs (like $\tau^+ \tau^-$, $W W^\ast$ etc.) can also be used in our analysis.  When the contributions from all other channels are added, exclusion or discovery limits derived in our analysis are expected to improve further.
The novelty of our study lies in the following points:
\begin{enumerate}
\item Although the role of jet kinematics in VBF has been studied in earlier works \cite{Englert:2012xt,Anderson:2013afp,Degrande:2016dqg}, we point out some hitherto unexplored kinematic features which can play crucial roles in differentiating the effects of additional operators from those of SM driven VBF, looking at the same final states. We illustrate this by using some higher dimensional operators which accentuate this difference.

\item The additional operators introduced here carry Lorentz structures that are distinct from the SM-induced one. The response of the event selection criteria are correspondingly different. We not only highlighted such difference but also attempted to utilise them in kinematic effects, by means of a correlated two-dimensional analysis 
between the geometric mean of the transverse momenta of the two forward jets and the rapidity difference between them, thereby uncovering new regions of phase-space to exploit the VBF production process.

\item Although it may have been noticed earlier, we underline the importance of bin-by-bin statistical significance in differential distributions, when it comes to distinguishing the additional dimension-6 operators, inducing new interactions of 4-point vertices of the form $qqV H, (V = W^\pm,Z)$.

\item We have duly estimated the next-to-leading order (NLO) QCD corrections to VBF Higgs production in presence of dimension-6 operators. The NLO QCD effect positively adds up to the LO rate without significantly affecting the overall shape of the kinematic distribution, albeitly, improving the distinctness from the SM in most of the bins.

\end{enumerate}

The outline of the paper is the following. In Section~\ref{sec2}, we review the basics of the effective field theory framework necessary for VBF Higgs production and discuss the motivation of this study followed by results of Monte Carlo study in Section~\ref{sec3}. In Section~\ref{sec4}, the effects of new dimension-6 interactions are analysed in the differential distributions of jet observables, by allowing one {\em non-zero} SMEFT coupling to vary at a time. We describe our phenomenological analyses in the VBF process by studying the sensitivity of these observables, along with the dependence on the LHC centre of mass energy; our main findings are summarised in Section~\ref{sec5}.

\section{Overview of SMEFT operators relevant for VBF Higgs signal}
\label{sec2}
\begin{figure}[!h]
	\centering
	\includegraphics[width=3.0cm,height=3.0cm]{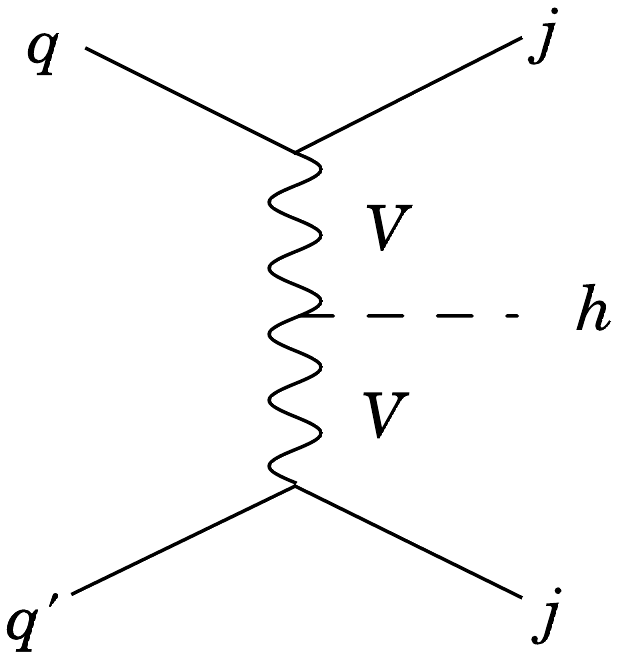}~
	\includegraphics[width=3.0cm,height=3.0cm]{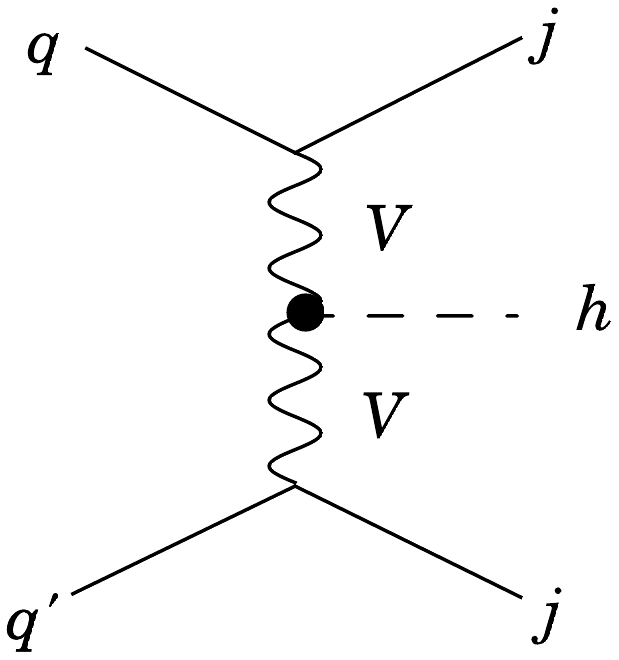}~
	\includegraphics[width=3.0cm,height=3.0cm]{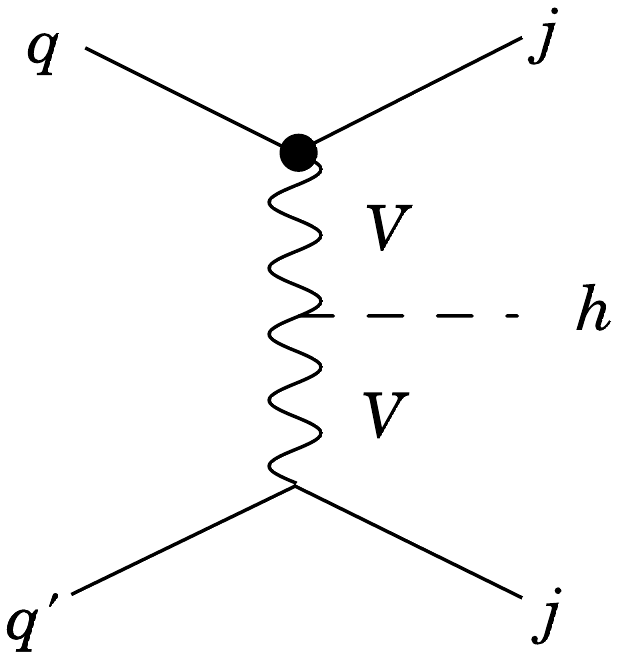}~
	\includegraphics[width=3.0cm,height=3.0cm]{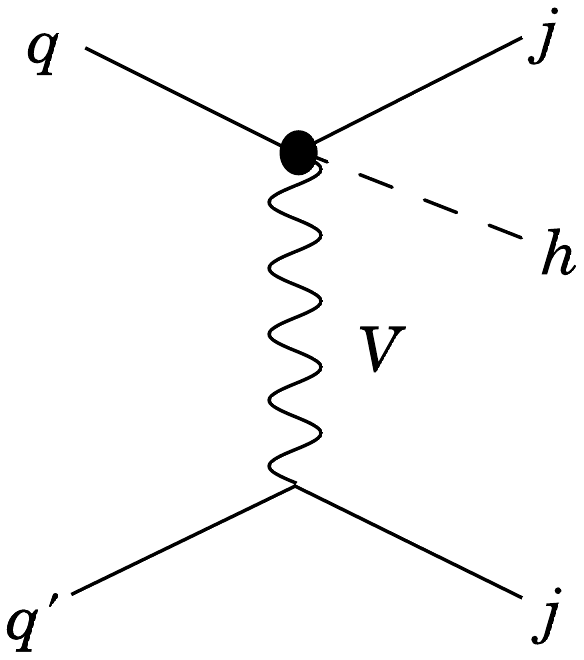}\\
	\hspace{2mm}(a) \hspace{27mm}(b) \hspace{26mm}(c) \hspace{27mm}(d)
	\caption{\label{tab:Feynmandiag} Representative Feynman diagrams contributing to Higgs production via Vector Boson Fusion topology. (a) is the SM process, while (b) and (c) processes involve the anomalous couplings of Higgs with gauge bosons and fermions respectively. (d) involves the  contact interaction between the Higgs, gauge boson and fermions denoted with a blob}	
\end{figure}

We focus  on the general set of dimension-6 gauge invariant operators which give modifications in the VBF Higgs production. To facilitate our  discussion, we present in Fig. \ref{tab:Feynmandiag}, the Feynman diagrams  which, by virtue of SM interactions and dimension-6 effective operators, contribute to VBF amplitude. Black dots on some of the vertices of diagrams (b), (c) and (d) 
stand for possible inclusion of one of the following higher dimensional operators \cite{Buchmuller,Grzadkowski:2010es} listed below.

\begin{itemize}
	\item The dimension-6 operators containing the SU(2) Higgs doublet $\vp$ and its derivatives:
	\begin{equation}
	\mathcal{O}_{H\Box}  =  (\vp^\dag \vp)\raisebox{-.5mm}{$\Box$}(\vp^\dag \vp);~~~
	\mathcal{O}_{H D} =  \left(\vp^\dag D^\mu\vp\right)^\star \left(\vp^\dag D_\mu\vp\right)  
	\end{equation}
	These operators modify the SM Higgs couplings to other particles by multiplicative factors without bringing in any new Lorentz structure. This amounts to a renormalization of the Higgs field. Here, the covariant derivative $D_\mu$ has the usual meaning and contain $SU(2)_L$, $U(1)_Y$  gauge couplings and bosons.
	\item The operators that induce fermion-fermion-gauge-Higgs ($qqVh$) or fermion-fermion-gauge ($qqV$)  interactions leading to the amplitudes like (c) or (d) in VBF are : 
	\begin{equation}
	\mathcal{O}_{H q}^{(1)} = (\vpj)(\bar q_p \gamma^\mu q_r);~~~
	\mathcal{O}_{H q}^{(3)} = (\vpjt)(\bar q_p \tau^I \gamma^\mu q_r) \nonumber
	\end{equation}
	\begin{equation}
	\mathcal{O}_{H u} = (\vpj)(\bar u_p \gamma^\mu u_r);~~~
	\mathcal{O}_{H d}  = (\vpj)(\bar d_p \gamma^\mu d_r) \nonumber
	\end{equation}
	\begin{equation}
	\mathcal{O}_{H u d} = i(\tvp^\dag D_\mu \vp)(\bar u_p \gamma^\mu d_r) 
	\end{equation}
	
		where we define 
		$\vpj= \vp^\dagger D_\mu \vp-(D_\mu \vp)^\dagger
		\vp$ and $\vpjt= \vp^\dagger \tau^I
		D_\mu \vp-(D_\mu\vp)^\dagger\tau^I \vp$.  We use the notation of $q$ for the quark doublet under $SU(2)_L$ and $u,d$ for the $SU(2)$ singlet quarks, $p$ and $r$ are generation indices.   
	
	Excepting the operator $O_{Hud}$ (which induces a right handed charged current), the VBF amplitudes constructed from any of these   four operators $O_{Hq}^{(3)},O_{Hq}^{(1)}, O_{Hu}$ and $O_{Hd}$ interfere with the SM amplitude.
	
	\item Another set of  operators can induce $qqVh$ or $qqV$ interactions. While their structure indicates that they are of magnetic dipole type and are different from the earlier ones, the presence of explicit dependence on the momentum of the gauge bosons in such operators due to the gauge field strengths,  will play a significant role in jet kinematics in which we are interested in this analysis. 

	\begin{equation}
	\mathcal{O}_{uW} = (\bar q_p \sigma^{\mu\nu} u_r) \tau^I \tvp\, W_{\mu\nu}^I;~~~
	\mathcal{O}_{dW} = (\bar q_p \sigma^{\mu\nu} d_r) \tau^I \vp\, W_{\mu\nu}^I \nonumber
	\end{equation}
	\begin{equation} 
	\mathcal{O}_{uB} = (\bar q_p \sigma^{\mu\nu} u_r) \tvp\, B_{\mu\nu};~~~
	\mathcal{O}_{dB} = (\bar q_p \sigma^{\mu\nu} d_r) \vp\, B_{\mu\nu} 
	\end{equation}

Neglecting fermion masses, the dipole operators connect fermions of  different helicities. Consequently, VBF amplitudes constructed out of such operators (Fig. \ref{tab:Feynmandiag} (c) and (d)) do not interfere with the SM amplitudes.
	\item 
	
	Finally, there exist anomalous $HVV$ interactions arising from the following set of dimension-6 operators \cite{Buchmuller,Grzadkowski:2010es}:
	
	\begin{equation}
\mathcal{O}_{HW} = \vp^\dag \vp\, W^I_{\mu\nu} W^{I\mu\nu};~~~
\mathcal{O}_{HB} = \vp^\dag \vp\, B_{\mu\nu} B^{\mu\nu};~~~
\mathcal{O}_{HWB} = \vp^\dag \tau^I \vp\, W^I_{\mu\nu} B^{\mu\nu} 
\end{equation}
	
These operators modify HVV couplings by introdcing new Lorentz strcuture in the Lagrangian. Consequently, new VBF amplitudes ((b) in Fig. \ref{tab:Feynmandiag}) arising from such interactions add to the SM amplitude. 
	
\end{itemize}

We present above a complete set of operators, which can modify the SM VBF Higgs signal. However, we illustrate our main points by using as samples the operators $\cO_{Hq}^{(1)}$, 
$\cO_{Hq}^{(3)}$ and $\cO_{uW}$ and our study focuses on the important effect of differently structured interactions in the selection efficiencies and kinematic observables. Moreover, the method developed here, is of general utility in studying all possible higher-dimensional operators. For instance, $\mathcal{O}_{fW}$ and $\mathcal{O}_{fB}$ operators differ in the cross section by a total factor $\rm tan^2 \theta_W$. Also, as can be seen from the discussion of Fig. \ref{tab:cxratio} in the following section, operators involving purely bosonic fields modify the VBF rates to a lesser extent than what fermionic operators do. We refer the reader to the discussion on Fig. \ref{tab:cxratio} in the next section, which will hopefully clarify why these three operators can be treated as representative.
 
 A comment about the numerical values of operator coefficients $\frac{C}{\Lambda^2}$, used in the following analysis is  relevant at this point. The ultimate guideline for the numerical values to be used, lies in the  available data from experiments. $\mathcal{O}_{H q}^{(1)}$ and  $\mathcal{O}_{H q}^{(3)}$ lead to Z and W couplings to fermions with Lorentz structure similar to the SM. The electroweak precision measurements at LEP-I and LEP-II lead to stringent constraints on the Wilson coefficients of these operators.  A global fit of electroweak observables to LEP data \cite{Rizzo:1994qz,ewpd wd} leads to bounds on effective vector and axial-vector coupling of a pair of fermions to a $Z$-boson, which can be, in turn, translated into  bounds  at 95\% C.L. on $\vert{ C_{Hq}^{(1,3)} \over \Lambda^2 }\vert$ which is 1.11 TeV$^{-2}$. Similar constraints cannot be imposed on $C_{uW}$ due to the chiral structure of $\cO_{uW}$. 
 However, one can think of imposing bounds  on  $C_{uW}$ by considering  its contribution to the anomalous magnetic moment of $u$-quark at tree level. But, for the light quarks, $(g-2)$  is hard to extract in a model independent way, is therefore subjected to large uncertainties \cite{Giudice:2012ms}.

Experimentally measured rates of nuclear beta decay as well as  leptonic and semi-leptonic decays of pions and kaons, also constrain the couplings in which we are interested. As for example, the allowed uncertainty  of the pion form factor implies  $\frac{|C_{Hq}^{(3)}|}{\Lambda^{2}} < 1.64$ TeV$^{-2}$ at 90\% C.L \cite{ParticleDataGroup:2020ssz}.

\begin{center}
\begin{table}[h!]
	\centering
	\scalebox{1.0}{
	\begin{tabular}{|c|c|}
		\cline{1-2} 
		\\[-1em]
		$Zh, h \to b\bar{b}$ &  95\% CL allowed range
		\\
		\hline
		${\cal O}_{Hq}^{(3)}$ & (-0.94, 0.41) \footnotesize{(TeV$^{-2}$)}
		\\
		\hline
		${\cal O}_{Hq}^{(1)}$  & (-0.72, 0.61) \footnotesize{(TeV$^{-2}$)}
		\\
		\hline
		${\cal O}_{uW}$  &(-0.68, 0.68) \footnotesize{(TeV$^{-2}$)}\\
		\hline
		\hline
		\\[-1em]
		$Wh, h \to b\bar{b}$  &  95\% CL allowed range
		\\
		\hline
		${\cal O}_{Hq}^{(3)}$ & (-1.11, 0.43) \footnotesize{(TeV$^{-2}$)}
		\\
		\hline
		${\cal O}_{uW}$  & (-0.82, 0.82) \footnotesize{(TeV$^{-2}$)}
		\\
		\hline
		\hline
		\\[-1em]
		$VBF, h \to \tau \bar{\tau}$ &  95\% CL allowed range
		\\
		\hline
		${\cal O}_{Hq}^{(1)}$  & (-5.4, 4.27) \footnotesize{(TeV$^{-2}$)}
		\\
		\hline
		${\cal O}_{Hq}^{(3)}$  & (-1.56, 3.7) \footnotesize{(TeV$^{-2}$)}		\\
		\hline
		${\cal O}_{uW}$  &(-2.06, 2.06) \footnotesize{(TeV$^{-2}$)}\\
		\hline
		
	\end{tabular}}
	\caption{ 95\% CL lower and upper limits (right column) on the relevant couplings $({C \over {\Lambda ^2} })$ of selected 
	dimension-6 operators (left column) obtained from comparing theoretical predictions with the combined ATLAS and CMS cross-section measurements \cite{ATLAS:2018kot, CMS:2018nsn, ATLAS:2018ynr, CMS:2017zyp} from 13 TeV run of the LHC.}
	 
	\label{tab:bf95}
\end{table}
\end{center}
The aforementioned operators are also subjected to the constraints imposed by the LHC data. In Table. \ref{tab:bf95}, we have listed the bounds taking one operator at a time, obtained by comparing the expected cross-section with experimental data from ATLAS and CMS collaborations \cite{ATLAS:2018kot, CMS:2018nsn, ATLAS:2018ynr, CMS:2017zyp}.

The most stringent limits on ${C_{Hq}^{(3)}}\over {\Lambda^{2}}$, ${C_{Hq}^{(1)}}\over {\Lambda^{2}}$ and ${C_{uW}} \over {\Lambda^{2}}$  arise from
 the measured cross-section, of associated production of Higgs with a vector boson (Vh production) because of higher accuracy in its measurement.

As already mentioned, $\mathcal{O}_{H\Box} $ and $\mathcal{O}_{H D}$ will renormalise the Higgs wave function and in turn, it will modify all the Higgs observables.  We have obtained bounds on these two operators by comparing the Higgs production rate via VBF with ATLAS data \cite{ATLAS:2018ynr}.  The 95\%  confidence intervals for these couplings are listed in the Table \ref{tab:bf96}. However, this set of operators will not give us Higgs couplings to other SM particles with new Lorentz structure and will only rescale the Higgs interactions. These contributions, thus, not changing the momentum structures of the vertices involved, are of limited interest to investigate the effects of such operators any further, in our study.
\begin{center}
\begin{table}[h]
	\centering
	\begin{tabular}{|c|c|}
		\cline{1-2}
		${\cal O}_{H\Box}$  &(-41.24, 8.28)  \footnotesize{(TeV$^{-2}$)}\\
		\hline
		${\cal O}_{HD}$  &(-17.57, 25.4) \footnotesize{(TeV$^{-2}$)}\\
		\hline
	\end{tabular}
	\caption{95\% CL allowed range (right column) on the relevant couplings $({C \over {\Lambda ^2} })$ of selected 
		dimension-6 operators (left column) obtained from comparing theoretical predictions with ATLAS
		 \cite{ATLAS:2018ynr} from 13 TeV run of the LHC.} 
	\label{tab:bf96}
\end{table}
\end{center}

Some theoretical considerations are also important and we now pay some attention to these. A guiding principle is a  good high energy behaviour of the scattering amplitudes on the inclusion of higher-dimensional operators involving such coupling.  The  scattering amplitudes constructed out of such effective operators must satisfy the unitarity bound, namely $ \vert {\cal R}e~a_0 \vert < 0.5$, where $a_0$ is the lowest partial wave amplitude. The violation of unitarity appears at energies of a few TeV for the values of Wilson coefficients allowed by the Higgs data, with the exact value depending upon the specific choice of operators and the process under consideration.  A simplified unitarity analysis $q \bar q \rightarrow Vh$ leads to an upper bound of $2.82 ~\rm TeV^{-2}$, $8.81 ~\rm TeV^{-2}$ and $1.41 ~\rm TeV^{-2}$ on $\vert{ C_{Hq}^{(3)} \over \Lambda^2 }\vert$, $\vert{ C_{Hq}^{(1)} \over \Lambda^2 }\vert$ and $\vert{ C_{uW} \over \Lambda^2 }\vert$ respectively, assuming an incoming parton inside a proton carries typically $1 ~\rm TeV$ of energy at the LHC.

 In the subsequent study, keeping such constraints in mind, the effective coupling strengths need to be consistent with the electroweak data, partial wave unitarity and the Higgs measurements. Within such constraints, we concentrate on the effective operators ${\cal O}_{Hq} ^{(1)}$, ${\cal O}_{Hq} ^{(3)}{~\rm and }$ ~${\cal O}_{uW}$ and examine their role in modifying the VBF rates. The following sections contain a description of our strategy and results.

\section{Collider Analysis}
\label{sec3}
We implemented the effective Lagrangian of SMEFT in ~{\tt{FeynRules}}~\cite{Alloul:2013bka}. {\tt MadGraph-5}\cite{Alwall:2014hca} has been used to generate parton-level events. The SM cross-sections have been estimated at the next-to-leading order (NLO) as implemented in {\tt Madgraph-5}. While generating events driven by new Physics, we assume that only LO SMEFT contribution is absorbed within the Wilson coefficient of the dimension-6 operators. We use {\tt NNPDF23NLO} parton distribution function \cite{pdf} with renormalisation and factorisation scale equal to half the Higgs mass ($\mu_R = \mu_F = \frac{m_H}{2}$).   We also checked that the results with other scale choice (\textit{viz}, at  scalar sum of transverse momentum of all final state products) do not differ by more than $5 \%$. The $p p \to h (\to \gamma \gamma)j j$ is generated at $ \sqrt{s} = 14$ TeV. The events are passed through {\tt Pythia8}~\cite{Sjostrand:2006za} for parton showering and hadronisation. We performed the detector simulation in {\tt Delphes}~\cite{deFavereau:2013fsa} for analysing the hadron level events. The jets are reconstructed by following the anti-k$_t$ algorithm using {\tt FastJet}~\cite{Cacciari:2011ma}.

We start by reminding the reader that  we would like to investigate how the VBF Higgs signal gets modified in the presence of  new dimension-6 operators. 
To present our case, we confine to the  di-photon decay channel of the Higgs boson. To validate our analysis, we applied the ATLAS \cite{Aaboud:2018xdt} cut-flow listed in Table \ref{cutflow}, described in \cite{Araz:2020zyh}, to the SM VBF Higgs production followed by its di-photon decay. As a preselection requirement, we selected events with photons with minimum transverse momentum of 25 GeV, within $|\eta| < 2.37$ and separated from each other with $\Delta R > 0.4$. The jets are reconstructed with radius parameter 0.4 with minimum 30 GeV transverse momentum and within $|\eta|< 4.5$.  Note that the cuts as listed in Table \ref{cutflow}  are  optimised to keep out the background to VBF in the form of Higgs production via gluon production along with two jets and also non-Higgs backgrounds.

\begin{table}[!h]
	\centering
	\begin{tabular}{|l|c|c|c|}
		\hline
		Cut & Ref \cite{Araz:2020zyh} efficiency  & Our MC efficiency & Events\\
		\hline
		Presel. & -& -& 15841\\
		$N_{jets}\geq2$ & 0.838 & 0.852 & 13497\\
		$N_{b-jet}=0$ & 0.968 & 0.997 & 13456\\
		$|\Delta\eta_{jj}| > 3$ & 0.756 & 0.798 & 10738 \\
		$\eta_{j_1}\cdot\eta_{j_2} < 0$  & 0.987 &0.982 & 10545 \\
		$M_{jj}>600$ (GeV)   & 0.796 & 0.846 & 8921\\
		$N_\gamma=2$   & 0.657 & 0.612 & 5459 \\
		$I^{R=0.4}_\gamma< 15\%$ & 0.998 & 0.996 & 5438 \\
		$\Delta R^{min}_{\gamma j}>1.5$ & 0.886 & 0.823 & 4475 \\
		$|\Delta\Phi_{\gamma\gamma,jj}|>1.5$  & 0.976 & 0.953 & 4265\\
		$\Delta\Phi_{j_1,j_2}<2$  & 0.610 & 0.583 & 2486 \\
		$122<M_{\gamma\gamma}<128$ (GeV) & 0.996 & 0.998 & 2481 \\
		$y^{min}_{j_{1,2}} < y_h < y^{max}_{j_{1,2}}$ &0.984 & 0.977 & 2424\\
		\hline
	\end{tabular}
	
	\caption{ \small  Relative efficiencies of each of the experimental cuts and the expected number of events for the signal cross-section in the
	 $H \rightarrow \gamma \gamma$ channel, for the Higgs production via Vector Boson Fusion, for an integrated luminosity of 3000 fb$^{-1}$ at the 14 TeV LHC.}
	\label{cutflow}
\end{table}

The net efficiency of our selection cuts  ($0.153$) agree with \cite{Araz:2020zyh} ($0.163$) rather closely. The purpose of this exercise 
is overall validation of our MC, so that we can extract the efficiencies of the same cut flow when dimension-6 SMEFT operators are included within our own setup.
 
As already mentioned in the previous section, several  operators can modify the SM VBF Higgs signal. The total rates and their ratios at two different centre of mass energies can uncover signatures of the new and anomalous couplings of Higgs to other SM particles. In general, the energy dependence of the rates can be sensitive to the effective operators. We study now,  the ratio of VBF Higgs cross-sections at 14 TeV and 13 TeV (at the LHC),  keeping one of the aforementioned dimension-6 operators non-zero at a time along with the SM.  Higgs production cross-section is expected to  be more sensitive to the energy of collision  in presence of any of the higher dimensional operators than in the situation when production dynamics is solely controlled by the SM.

\begin{figure}[H]
	\centering
	\subfloat{
		\begin{tabular}{cc}
			\includegraphics[width=7.5cm,height=6.4cm]{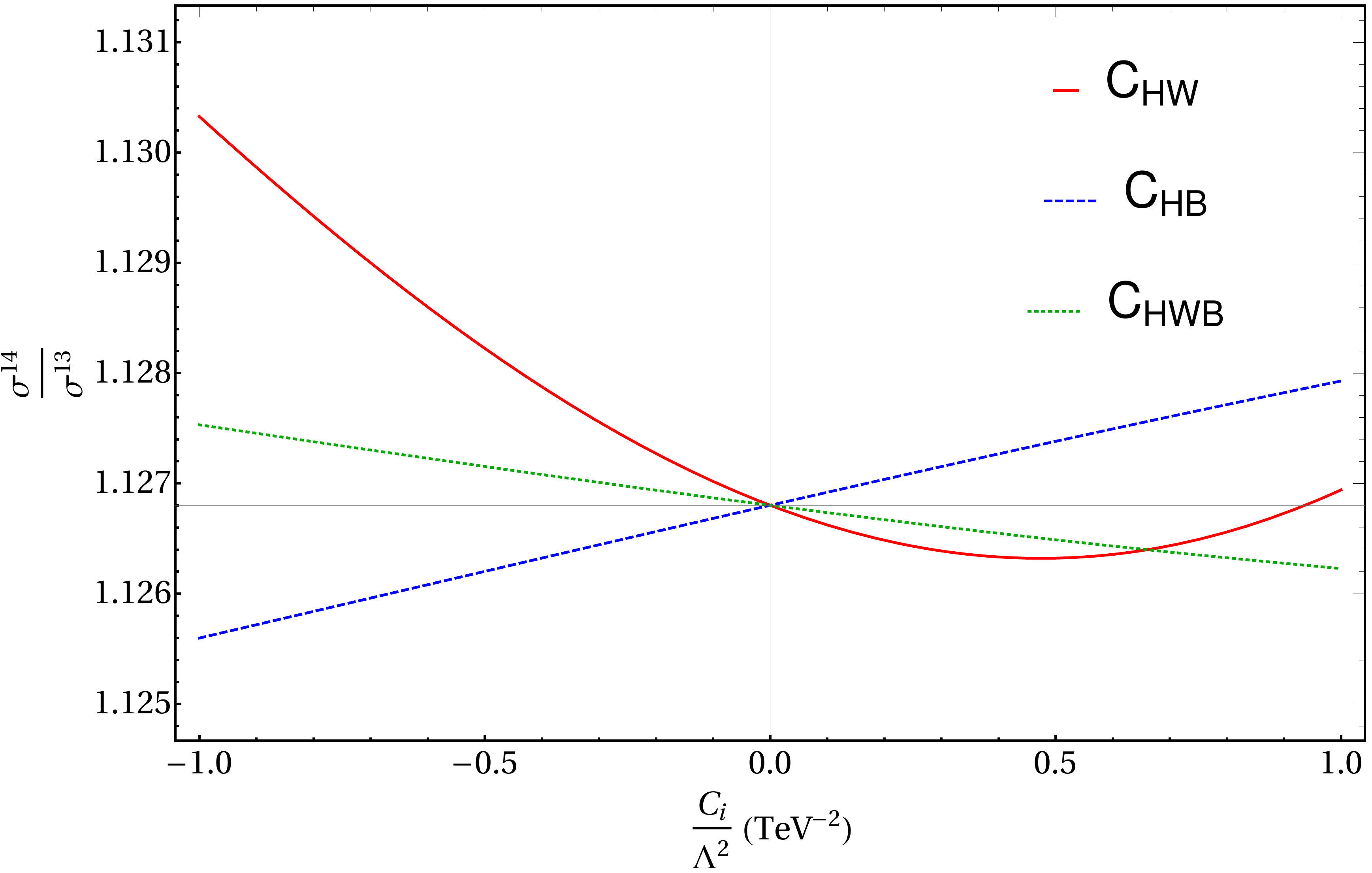}&
			\includegraphics[width=7.5cm,height=6.4cm]{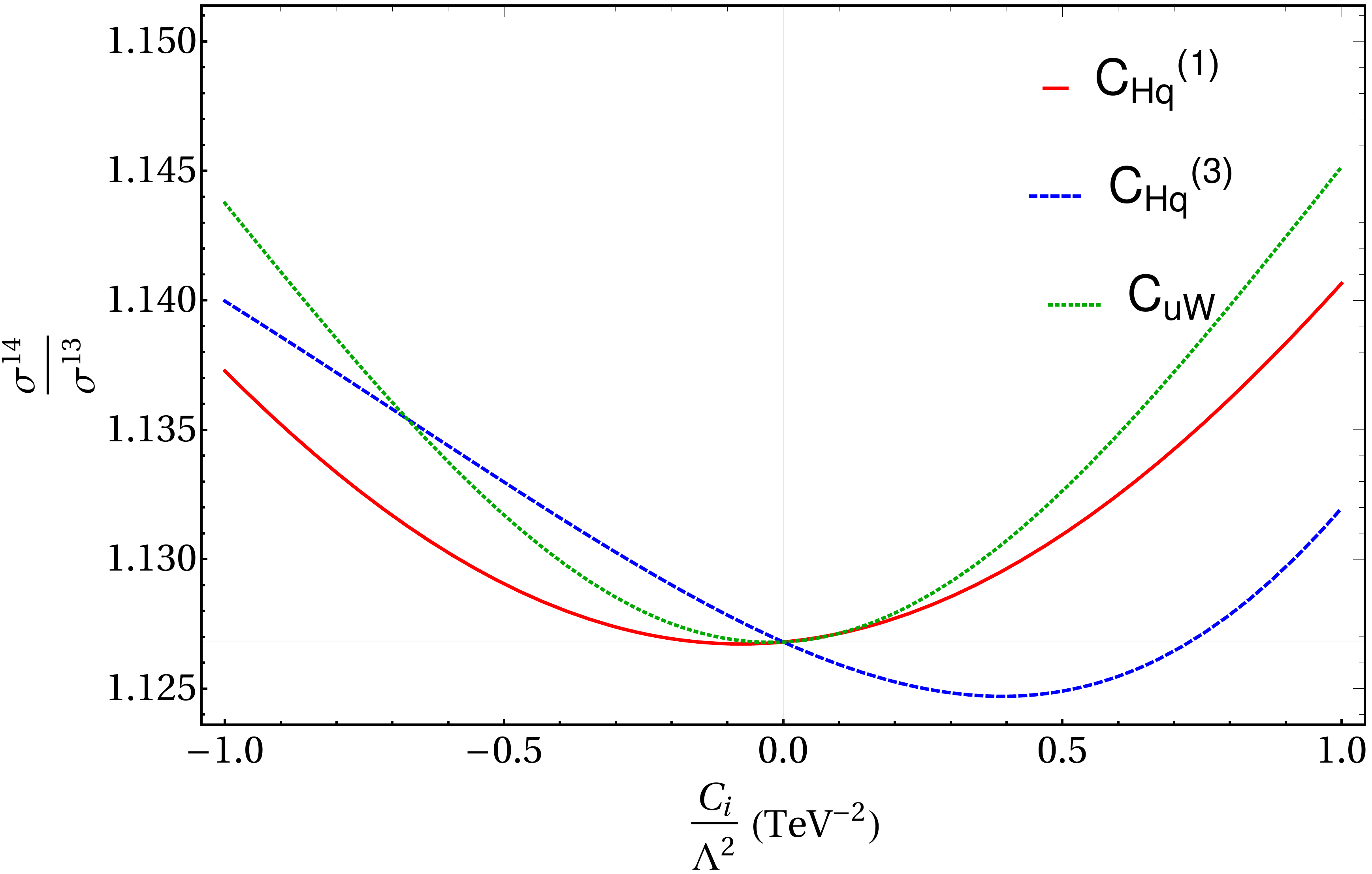}\\
			\hspace{4mm}(a)&\hspace{4mm}(b)
		\end{tabular}}
		\caption{Variation of  ratio of cross-sections at 14 TeV and 13 TeV for $pp \to hjj ( \to \gamma \gamma jj)$ with (a) $C_{HW}, C_{HB}, C_{HWB}$   and (b) $C_{Hq}^{(1)}, C_{Hq}^{(3)}, C_{uW}$ }
		\label{tab:cxratio}
\end{figure}

 The variation of the  ratios of the cross-section at LHC centre of mass energy at 14 TeV to 13 TeV   are presented in Fig. \ref{tab:cxratio}(a) and (b) respectively with varying Wilson coefficient values of various dimension-6 operators.  The effects of the different operators to show up to different degrees in such a ratio.  A grey line (corresponding to the value of the ratio of 1.268) parallel to x-axis in both the plots represents the SM case, where the relative enhancement of the cross-section is mainly due to parton flux evaluated at 14 TeV  {\em vis-a-vis} that at 13 TeV.
 
 A look into  Fig. \ref{tab:cxratio}(a) reveals that the bosonic operators ($\mathcal{O}_{HW}$, $\mathcal{O}_{HB}$ and $\mathcal{O}_{HWB}$) has a mild effect on the VBF Higgs cross-section in contributing to the SM. For values of $C_i/\Lambda^2$ varying in the range $[-1:0]$ TeV$^{-2}$, the deviation of the ratio from its SM value is the least for $\mathcal{O}_{H q}^{(1)}$ in comparison to other two fermionic operators as shown in  Fig. \ref{tab:cxratio}(b). In the same range of values of $C_i/\Lambda^2$,   $\mathcal{O}_{HW}$ driven ratio would deviate the most, from its SM value, in comparison to two other bosonic operators, $\mathcal{O}_{HB}$  and $\mathcal{O}_{HWB}$. However, the $C_{HW}$-driven ratio is less pronounced everywhere in this range of $C_i/\Lambda^2$, than the $C_{Hq}^{(1)}$-driven ratio. The ratio changes by  0.92\% for $C_{Hq}^{(1)}$ whereas for $C_{HW}$, it changes by 0.32\% for $ C_i/\Lambda^2$ in the range [-1:0] TeV$^{-2}$. Similarly, for positive values of Wilson coefficients, the 
 $C_{HB}$-driven ratio deviates the most from the SM value  and in the same range, the least deviation occurs due to   $\mathcal{O}_{H q}^{(3)}$. However, the latter is greater in magnitude than the $C_{HB}$-driven ratio in the range [0:1] TeV$^{-2}$. In a nutshell, for a given value of $C\over \Lambda^2$, any of the dimension-6 operators involving interaction of quarks, gauge and Higgs can modify the SM cross-section more than any of the operators involving anomalous coupling of bosons and Higgs only. Henceforth, we will only investigate the effects of $\mathcal{O}_{H q}^{(1)}$, $\mathcal{O}_{H q}^{(3)}$ and $\mathcal{O}_{uW}$, from the aforementioned groups of dimension-6 operators as they modify the VBF cross-section the most. 
 
\begin{figure}[H]
	\centering
	\subfloat{
		\begin{tabular}{cc}
			\includegraphics[width=7.5cm,height=6.4cm]{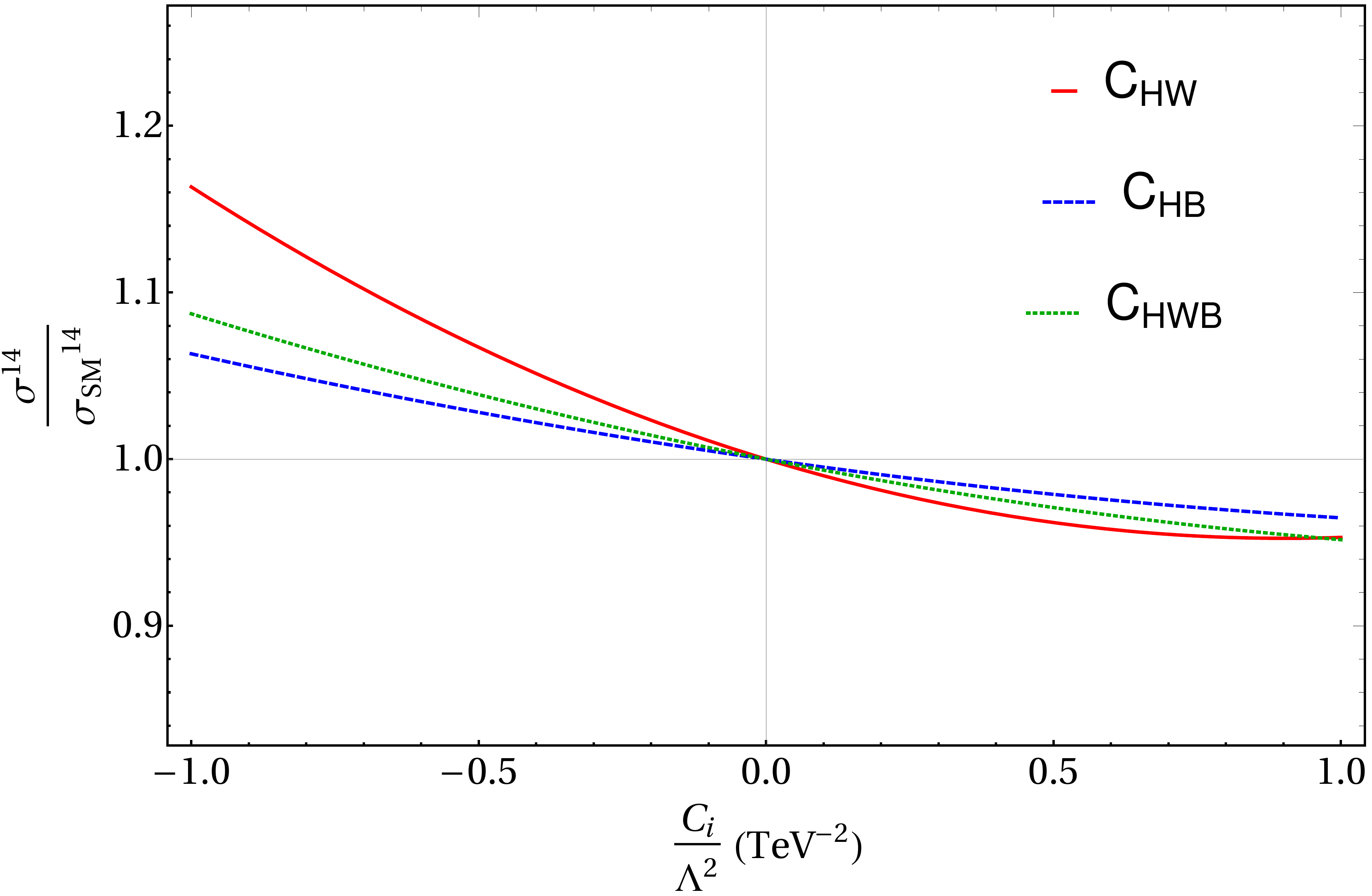}&
			\includegraphics[width=7.5cm,height=6.4cm]{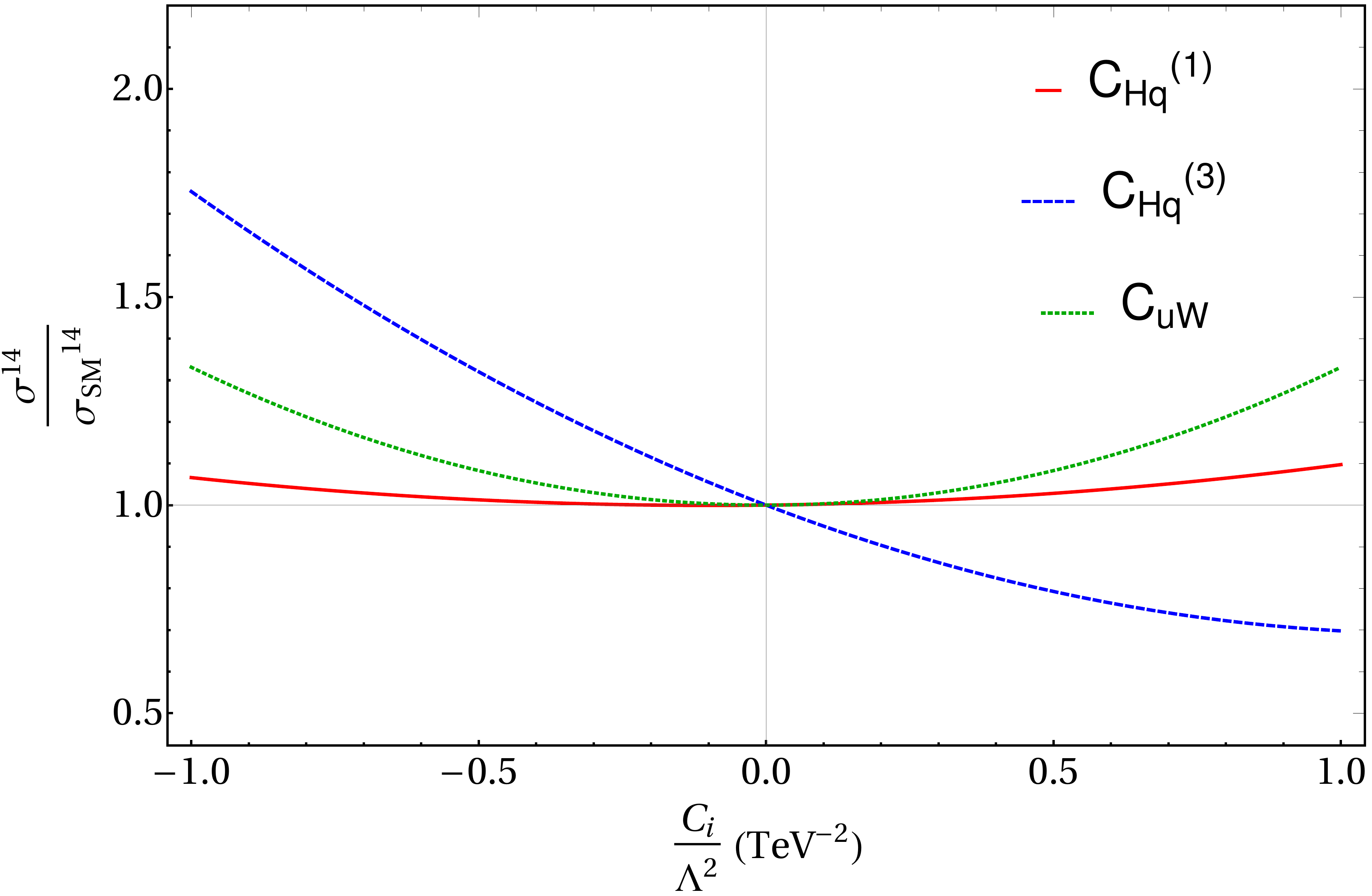}\\
			\hspace{4mm}(a)&\hspace{4mm}(b)
		\end{tabular}}
		\caption{Variation of  ratio of cross-sections of VBF Higgs production in presence of Dimension-6 operators to the SM cross-section of the same at 14 TeV   with (a) $C_{HW}, C_{HB}, C_{HWB}$   and (b) $C_{Hq}^{(1)}, C_{Hq}^{(3)}, C_{uW}$ }
		\label{new_ratio}
\end{figure}

 We have also estimated the strength of VBF Higgs production cross-section for different values of effective couplings normalised to the  SM cross-section.   Our finding on which operator gives the maximum effect remains unchanged. In Fig. \ref{new_ratio}, the variation of cross-section ratios have been presented. One can easily see from Fig. \ref{new_ratio} that in case of the operators involving two fermions, a gauge boson and a Higgs (right panel) difference of the ratios from unity, are more pronounced than the cases involving the bosonic operators (left panel).
 
 The enhancement of cross-section in case of dimension-6 operators involving fermions can be accounted by the absence of an extra propagator which is present in the SM-like VBF processes involving the bosonic operators. Most of the high-energy contribution due to these operators have an amplitude that is distinct from the SM contribution because of a quadratic growth with respect to the Mandelstam variable $t$.

A comment is in order here. Two of our operators ($\mathcal{O}_{H q}^{(1)}, ~\mathcal{O}_{H q}^{(3)}$) do interfere with the SM contributions while $\mathcal{O}_{uW}$ does not. One thus expects new contributions proportional to both $1 \over \Lambda^2$ and $1 \over \Lambda^4$. While it is expected to a general study to include all the contributions in a given order (${1 \over \Lambda^2}, {1 \over \Lambda^4}$), we are illustrating our points in the context of a simplified scenario {\it when one new operator arises at a time}. The discussion otherwise becomes so non-transparent and unwieldy that our main emphasis, namely the influence of the differently structured operators on jet-kinematics is lost. In the same spirit, when we are considering contributions upto quadratic order, $(\mathcal{O}\sim {1\over \Lambda^4})$ we have neglected the existence of additional dimension-8 operators which could participate at the same order. We reiterate that such simple-mindedness gives us the chance to explore the physical content of each operators.

The dependence of the efficiency (of the cuts)\footnote{The efficiency as a function of the parameters $C_{i}$ is defined as $\epsilon_{\gamma \gamma+ 2{\rm -jets}} (C_i) = \frac{\left[\sigma(pp \rightarrow H q q \rightarrow \gamma \gamma j j)  \right]_{\rm After ~Cuts}^{\rm VBF }} {\left[\sigma(pp \rightarrow H q q \rightarrow \gamma \gamma j j)  \right]_{\rm Before ~Cuts}^{\rm VBF }}.$}  on a particular Wilson-coefficient, will largely be controlled by interference with SM amplitudes. This feature will be more clearly revealed if the efficiency of the cuts (listed in Table \ref{cutflow}) are parametrised as a function of the Wilson coefficients.  

Efficiencies are found  to be second order polynomial functions of the Wilson coefficients. We calculate the total cross-section in the presence of higher dimensional operators at LO, in corroborating the fact that any higher order correction can affect the dependence of the cross-section on the coefficient of dimension-6 operators non-trivially \cite{Nason:2009ai}. Each power of coefficient $C_i$ is suppressed by $\Lambda^2$. For various  choices of this cut-off scale $\Lambda$, the coefficients will be scaled according to the  power of $C_i$ involved. We have  neglected any contribution from dimension-8 operators to the signal. The error due to this truncation at dimension-6 level cannot be estimated in a model independent way \cite{Contino:2016jqw}.

The efficiency corresponding to each of the  coefficients $C_i$ can be expressed as :

\begin{eqnarray}
\epsilon_{\gamma \gamma + 2-{\rm jets} (VBF)} (C_{Hq}^{(1)}) & =  &\dfrac{(0.6553 + \frac{C_{Hq}^{(1)}}{\Lambda^2} \;0.02327  +  (\frac{C_{Hq}^{(1)}}{\Lambda^2})^2 \;0.04144)}{(4.26 + \frac{C_{Hq}^{(1)}}{\Lambda^2} 0.1905  +  (\frac{C_{Hq}^{(1)}}{\Lambda^2})^2~0.3462)}, \nonumber\\
\epsilon_{\gamma \gamma + 2-{\rm jets} (VBF)} (C_{Hq}^{(3)}) & = & \dfrac{(0.6553 - \frac{C_{Hq}^{(3)}}{\Lambda^2} 0.3808  +  (\frac{C_{Hq}^{(3)}}{\Lambda^2})^2 0.16604)}{(4.26 - \frac{C_{Hq}^{(3)}}{\Lambda^2} 2.373 +  (\frac{C_{Hq}^{(3)}}{\Lambda^2})^2 1.347)}, \nonumber \\
\epsilon_{\gamma \gamma + 2-{\rm jets} (VBF)} (C_{uW}) & = & \dfrac{(0.6553 +  (\frac{C_{uW}}{\Lambda^2})^2 0.27)}{(4.26 + (\frac{C_{uW}}{\Lambda^2})^2 2.76)}
\label{eff-bb-gaga}
\end{eqnarray}
where the coefficients $C_i\over {\Lambda^2}$  are in units of $\rm TeV^{-2}$.

Some clarification is in order at this stage. The quantities of the form $C\over {\Lambda^2}$ are intrinsically 
dimensionful; therefore  $\vert {C\over {\Lambda^2} } \vert \simeq 1$ in this unit does not
necessarily imply a breakdown of the EFT expansion. It means that the ultraviolet (UV)
completion of our effective Lagrangian is a strongly coupled theory for large
$\Lambda$, while for relatively small $\Lambda$, it entails a weak UV completion.
Furthermore, for example in the first line of Eqn. \ref{eff-bb-gaga}, the term proportional to the square of 
 $C^{(1)}_{Hq} \over {\Lambda^2}$ exceeds the interference term in this limit, which may raise questions about the validity
of the EFT expansion. However, such a doubt is dispelled, so long as the interference of the
subsequent dimension-8 terms with the SM amplitude remains smaller than the aforementioned quadratic
terms. It may be legitimately expected that such an interference term is indeed smaller due to the
occurrence of $g_{SM}^2$ in it, so long as the energy dependence of the Wilson coefficients in the
dimension-8 terms are similar to those in the dimension-6 square terms \cite{Contino:2016jqw,LHCHiggsCrossSectionWorkingGroup:2016ypw}. This hopefully conveys an idea about the scope of Eqn. \ref{eff-bb-gaga}.

\begin{figure}[!h]
	\centering
	\includegraphics[width=11.0cm,height=9.0cm]{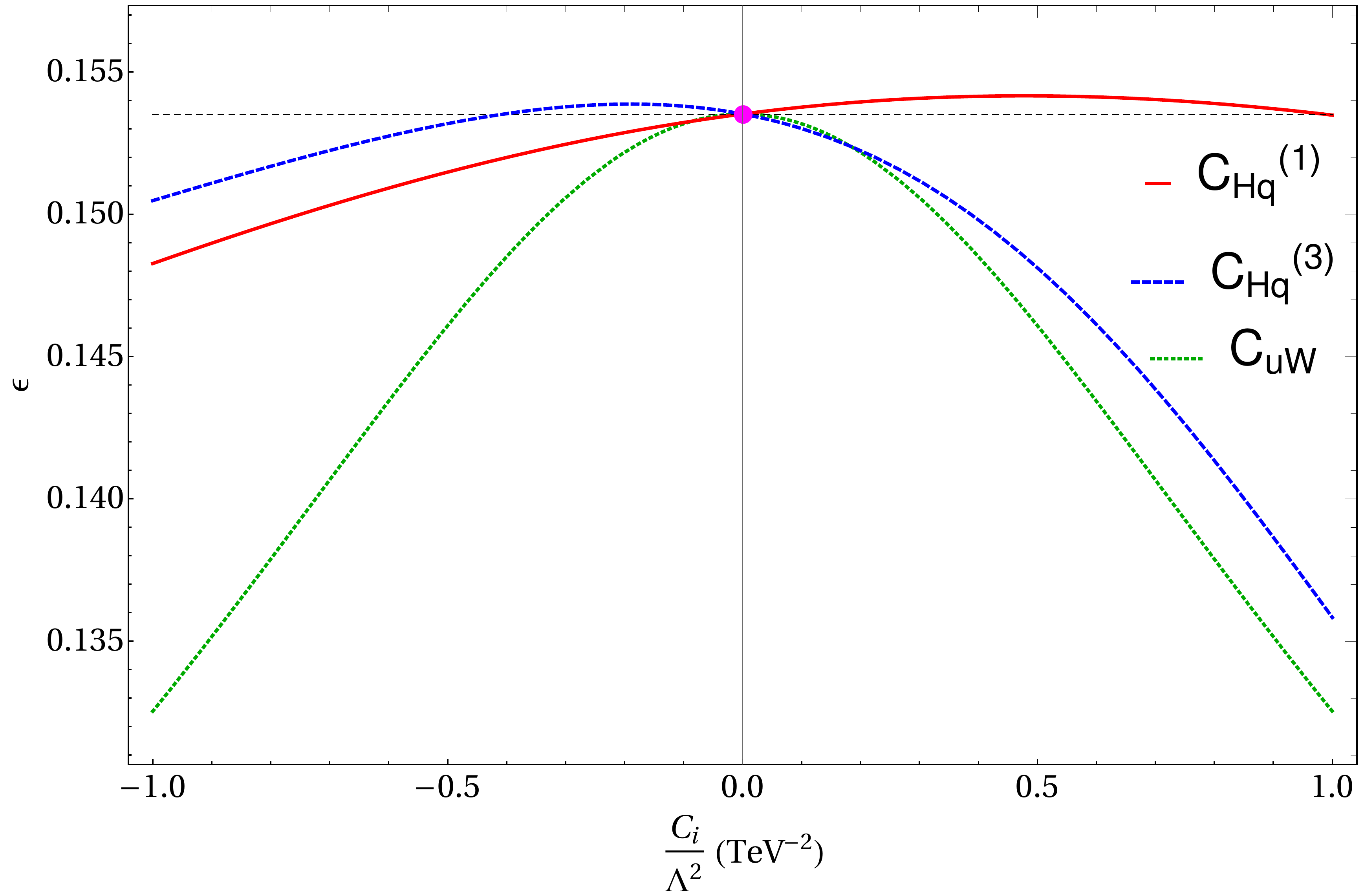}~
	\caption{\label{tab:Efficiency} Efficiency of VBF cuts ($\epsilon$) as a function of $\frac{C_i}{\Lambda^{2}}$ in the $h \to \gamma \gamma$ channel at 14 TeV}
	\label{efficiency}
\end{figure} 

In Fig. \ref{efficiency}, we show how efficiency depends on the Wilson coefficients, $C_{Hq}^{(1)}$, $C_{Hq}^{(3)}$ and $C_{uW} $.     A  small range of $C_i$ ( $ \vert \frac{C_{i}}{\Lambda ^2} \vert < 1 ~\rm TeV^{-2} $) consistent with unitarity and LEP
limits have been used in these plots for illustration.   We recover the SM efficiency $\epsilon_{SM} \simeq 0.153$, for zero values of these couplings. A symmetric nature of the green curve corresponding to $C_{uW}$,  reveals the non-interfering nature of corresponding operator, $\cO_{uW}$ with the SM. Here, the $\mathcal{O}(\frac{1}{\Lambda^{4}})$ term in the cross-section contains the leading BSM effect. The other two, namely, $\mathcal{O}_{H q}^{(1)}$ and  $\mathcal{O}_{H q}^{(3)}$, do interfere with SM amplitudes,  albeit differently, which can be easily realised by presence of a positive (negative) sign in front of  linear (in $C_i$) terms in the expressions of efficiencies in  Eqn. \ref{eff-bb-gaga}. The linear dependence of the cross-section on  $C_{Hq}^{(3)}$ is more sensitive than that on $C_{Hq}^{(1)}$, however, including the full amplitude-squared guarantees it to be positive-definite and would facilitate in distinguishing it from the SM. A naive scaling of the SM coupling alone by a multiplicative factor does not change of efficiency of cuts.

The efficiency as a function of $C_i \over \Lambda^2$ alone gives us limited information about how much an operator can modify the VBF Higgs cross-section from its SM value. The acceptance of these operators are not the same in the same regions of phase space and we must take into account the total and differential cross-section along with the efficiency to predict the above modification. 

We are now ready to investigate how the chosen dimension-6 operators may affect the various kinematic distributions. Our emphasis will be on those kinematic variables which can be constructed out of the four-momenta of tagging jets.  Our aim is to extract maximum information from the jet-observables without looking at the Higgs boson decay products so that the strategy followed in this article could be used for any other decay channels of Higgs boson produced via VBF mechanism. 

In the following, we will illustrate few of them in which we find the new physics effect is prominent. A value of ${C \over \Lambda^2} = 0.3 ~\rm TeV^{-2}$ has been used in these distributions.

\begin{figure}[H]
	\centering
	\subfloat{
		\begin{tabular}{cc}	
			\includegraphics[width=8.0cm,height=7.0cm]{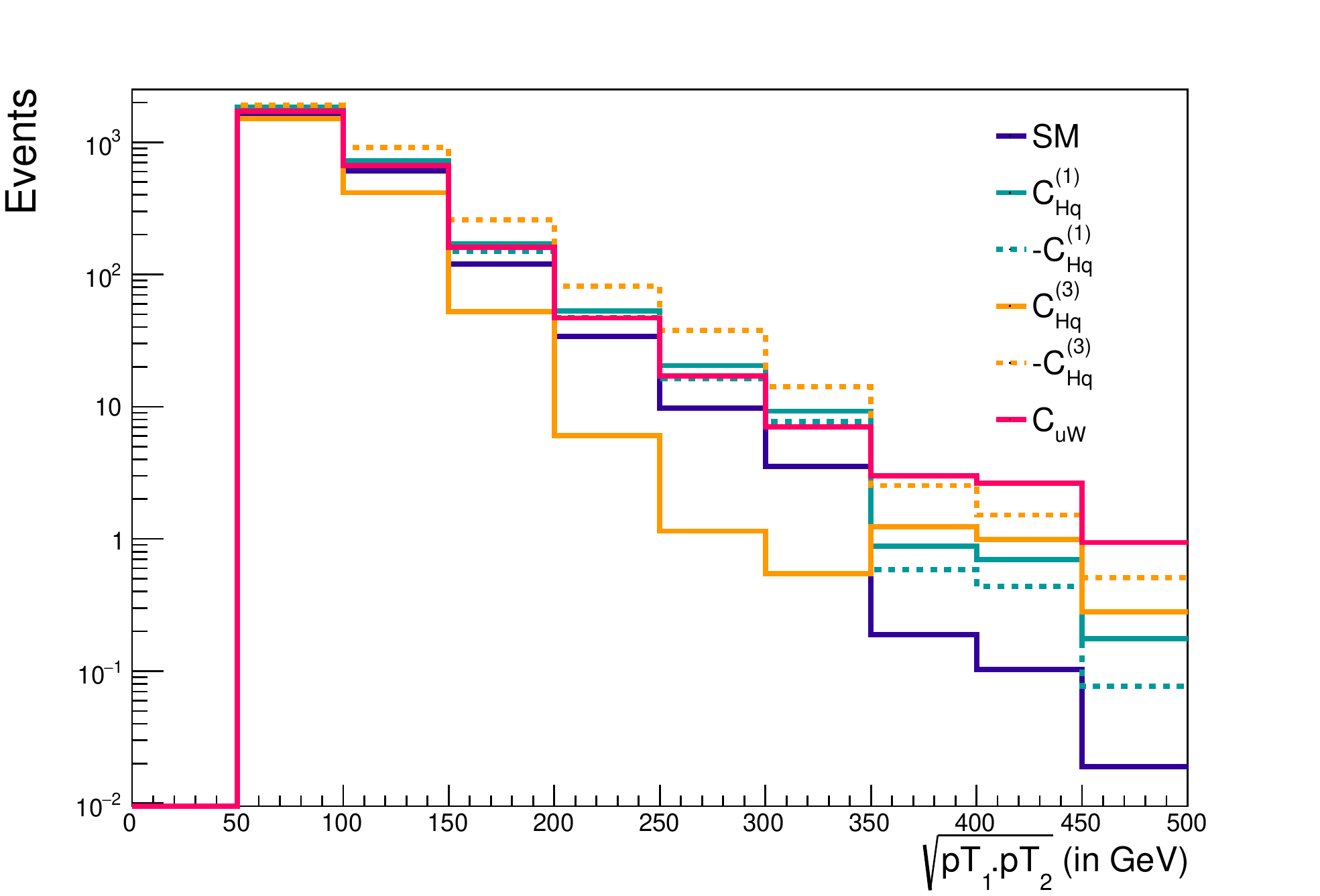}&
			\includegraphics[width=8.0cm,height=7.0cm]{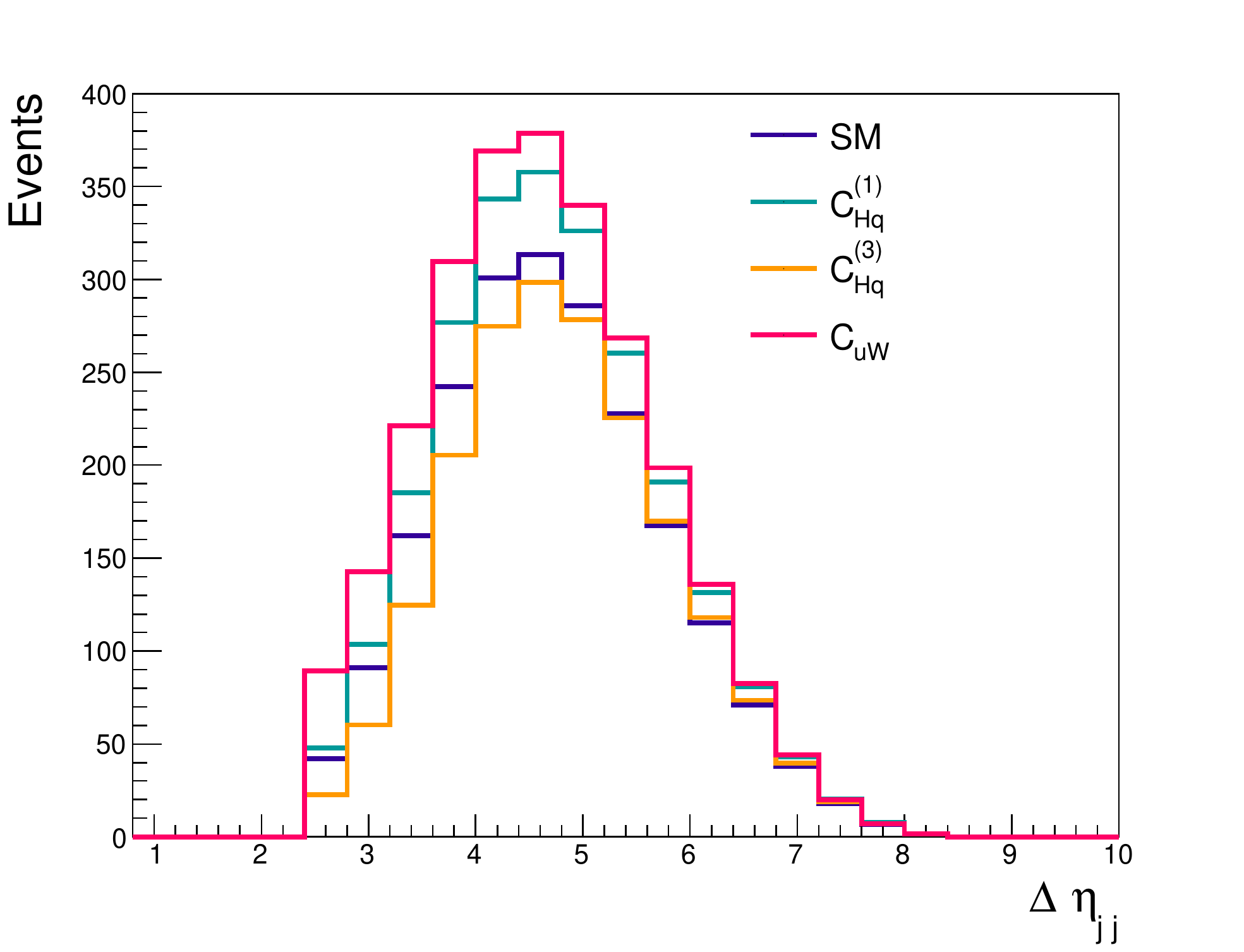}\\
			\hspace{4mm}(a)&\hspace{8mm}(b)
		\end{tabular}}
	\caption{(a) $p_{T12}$, geometric mean of leading jet $p_{T}$s and  (b) $\Delta \eta_{jj}$, rapidity separation between leading jet pair distributions in SM and SMEFT. The differential distribution of events are presented with 3000 $fb^{-1}$ data at 14 TeV LHC. For the EFT predictions, we have chosen $\frac{C_{Hq}^{(1)}}{\Lambda^2} = 0.3~\rm TeV^{-2}$ (green, solid), $\frac{C_{Hq}^{(3)}}{\Lambda^2} = 0.3~\rm TeV^{-2}$ (yellow, solid) and $\frac{C_{uW}}{\Lambda^2} = 0.3~\rm TeV^{-2}$ (red, solid). The dashed lines in (a) represent the same values of the respective Wilson coefficients but with negative signs. }
	\label{tab:vbfkinematicdistrbn}	
\end{figure}

The foremost is the geometric mean of $p_T$ of forwards jets, $ p_{T12} ~(\equiv \sqrt{ p_{T1} p_{T2}})$ distribution\footnote{In presence of dimension-6 operators, both the forward jets have higher $p_T$ compared to the SM case. To capture this enhancement in one distribution, we choose  $p_{T12}$ instead  of individual $p_T$ of the tagging jets.} as shown in Fig. \ref{tab:vbfkinematicdistrbn}(a). Sensitivity to dimension-6 operators are more pronounced at the tails of  $p_{T12}$ of the jets when we compare them with the SM. A few important aspects of this distribution are worth noting.

\begin{itemize}
\item Both $\mathcal{O}_{H q}^{(1)}$ and  $\mathcal{O}_{H q}^{(3)}$ interfere with the SM amplitude. The solid lines correspond to positive values of Wilson coefficients. The corresponding dashed lines correspond to negative values of the coefficients. The former (green solid line) enhances the event population steadily with increasing bins of $p_{T12}$ with respect to the SM (blue line). However, the absolute value for the excess events from the SM keeps on decreasing.

\item The operator  $\mathcal{O}_{H q}^{(3)}$ (yellow line) has destructive interference with SM, which is evident from the suppression of number of events in moderately low  $p_{T12}$ bins ($ < 350$ GeV) in comparison to the SM, while for  $p_{T12}$ bins ($ > 350$ GeV) there is substantial contribution due to the quadratic contribution of this operator in this region of phase space, relative to the destructive interference. The sign of the Wilson coefficient $C_{Hq}^{(3)}$ does not affect the hardness of the distribution in these regions as is noticeable from the distribution corresponding to negative value of the EFT coupling.

\item $\mathcal{O}_{uW}$ does not interfere with the SM. Presence of an explicit momentum (of the weak gauge boson) in the coupling  helps in producing higher number of jets with high $p_T$. 

\item If a negative departure from the SM is observed in the measured $p_{T12}$ (or $p_T$) distribution(s) of the leading forward jets in VBF events, that not only points towards a new interaction but also ensures  a specific form of new physics {\em e.g.} either  $\mathcal{O}_{H q}^{(1)}$ or  $\mathcal{O}_{H q}^{(3)}$. 

\end{itemize}

\begin{figure}[h!]
	\centering
	\subfloat{
		\begin{tabular}{cc}
				\includegraphics[width=7.5cm,height=6.0cm]{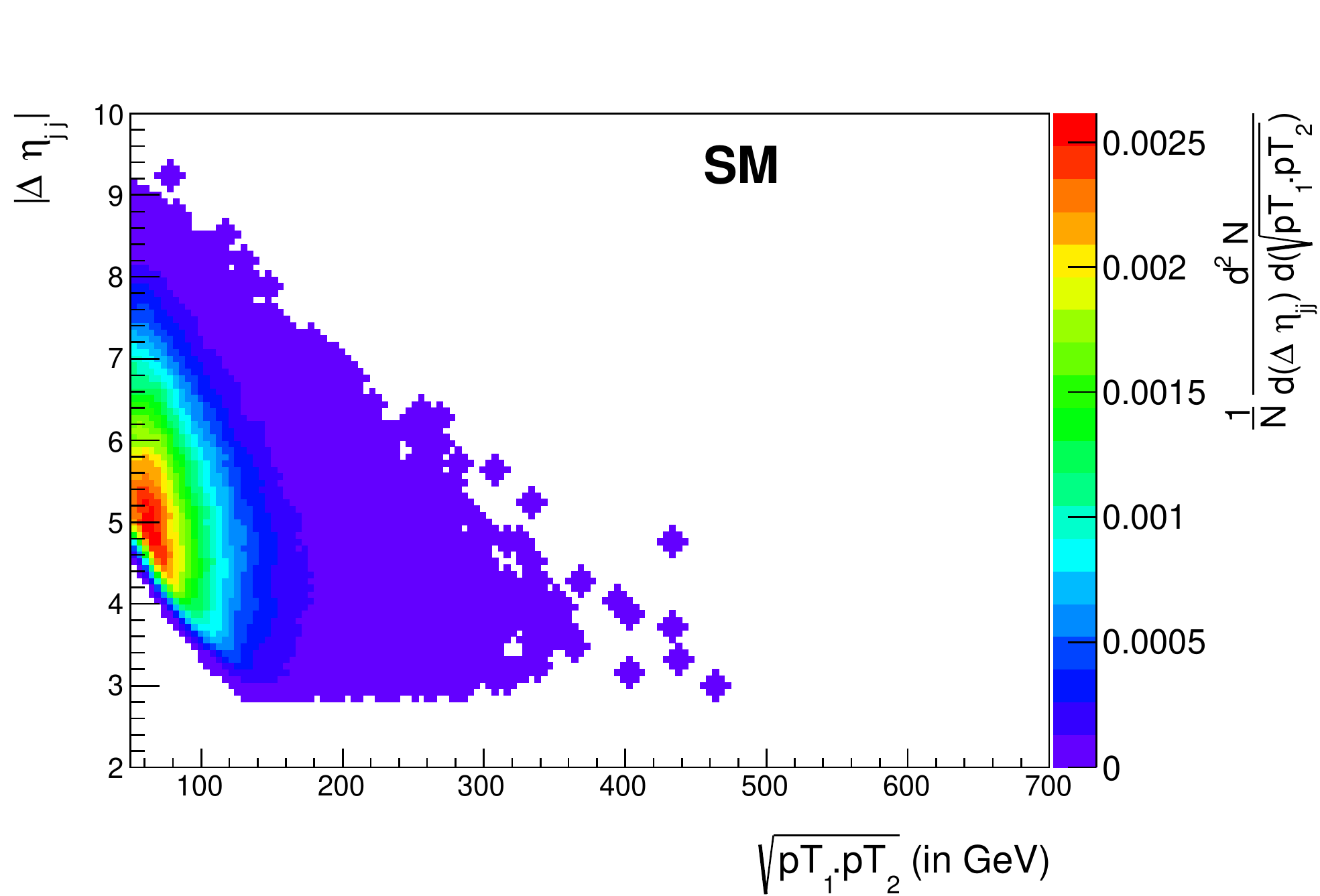}&
				\includegraphics[width=7.5cm,height=6.0cm]{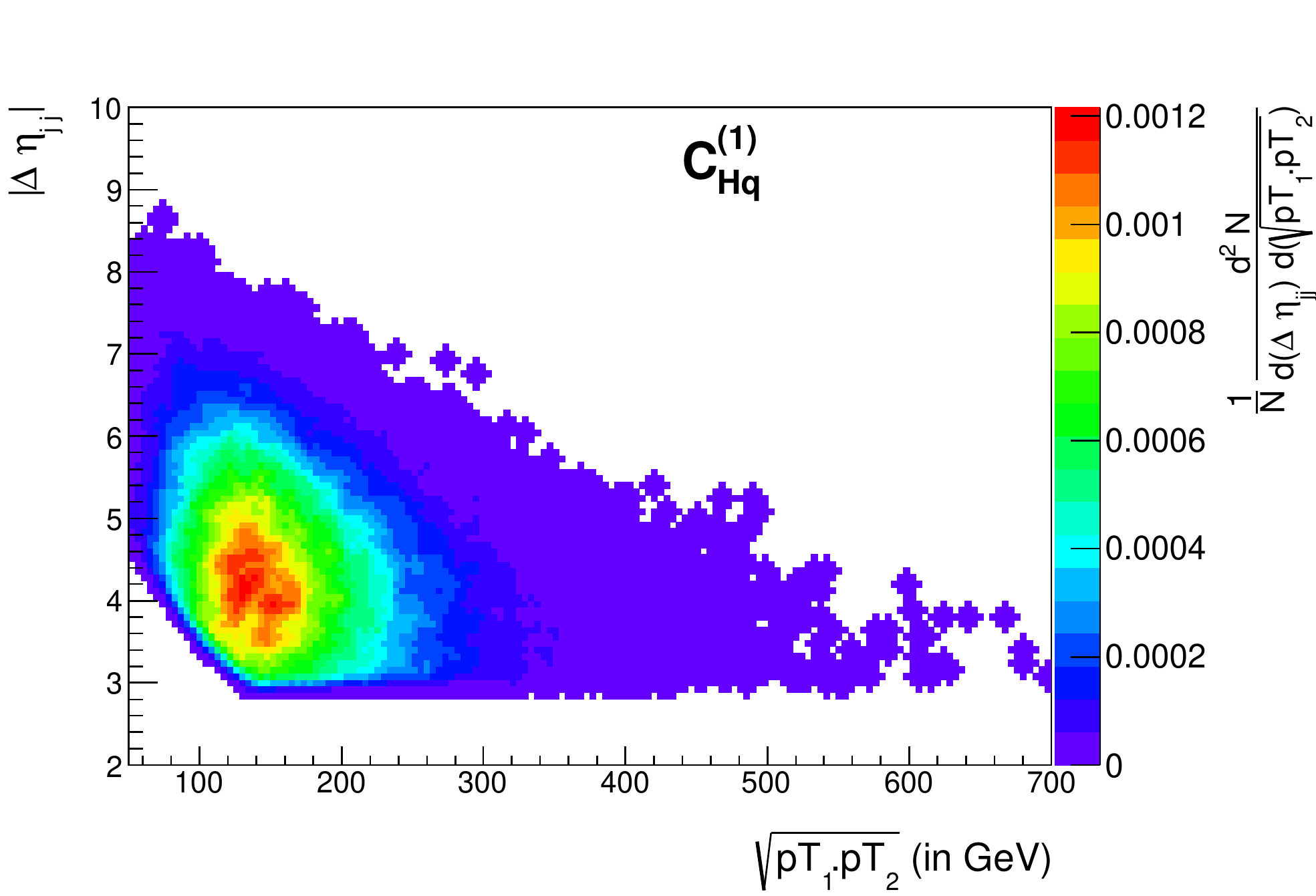}  \\
				\hspace{4mm}(a)&\hspace{8mm}(b)\\
				\vspace{1mm}\\
				\includegraphics[width=7.5cm,height=6.0cm]{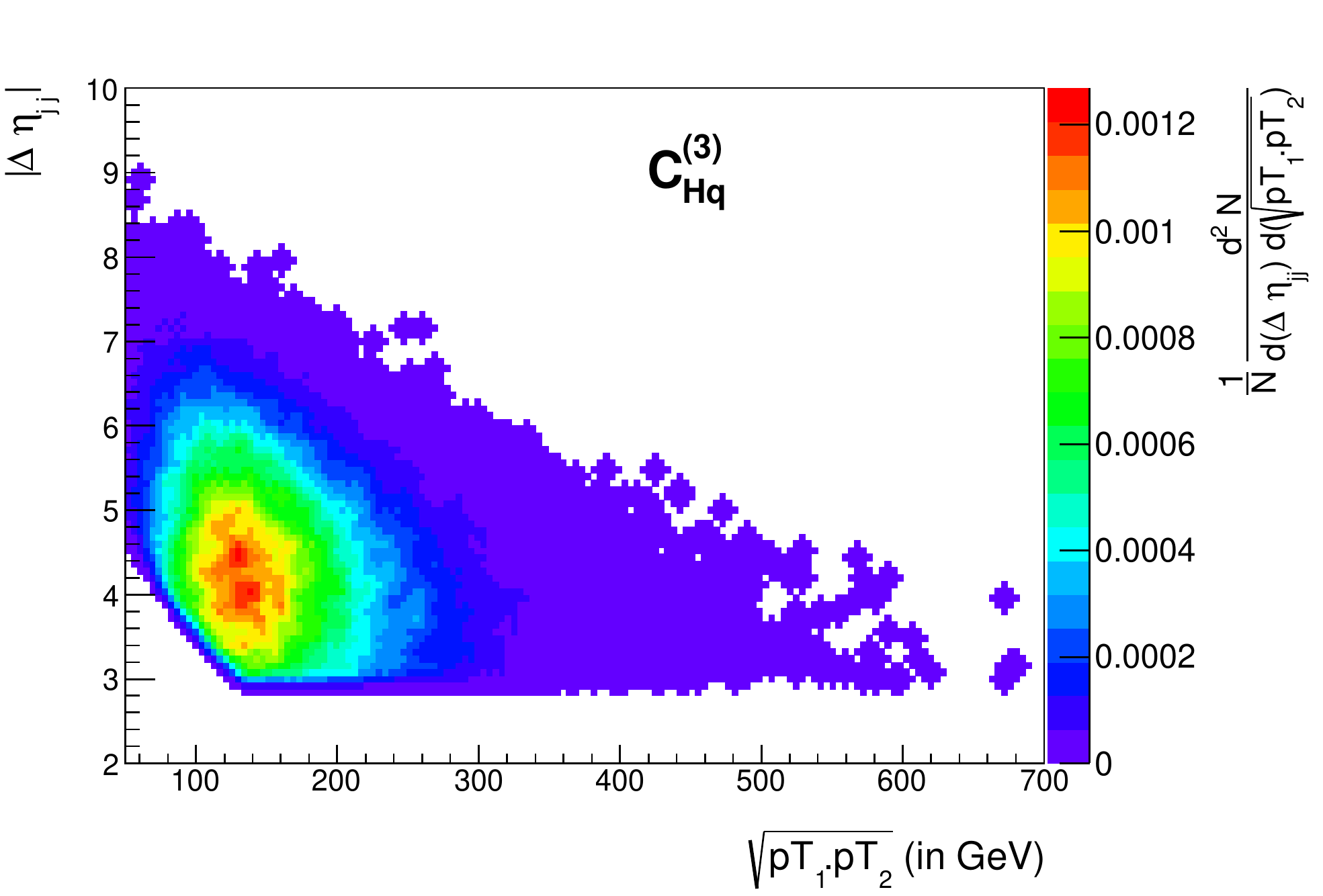}&
				\includegraphics[width=7.5cm,height=6.0cm]{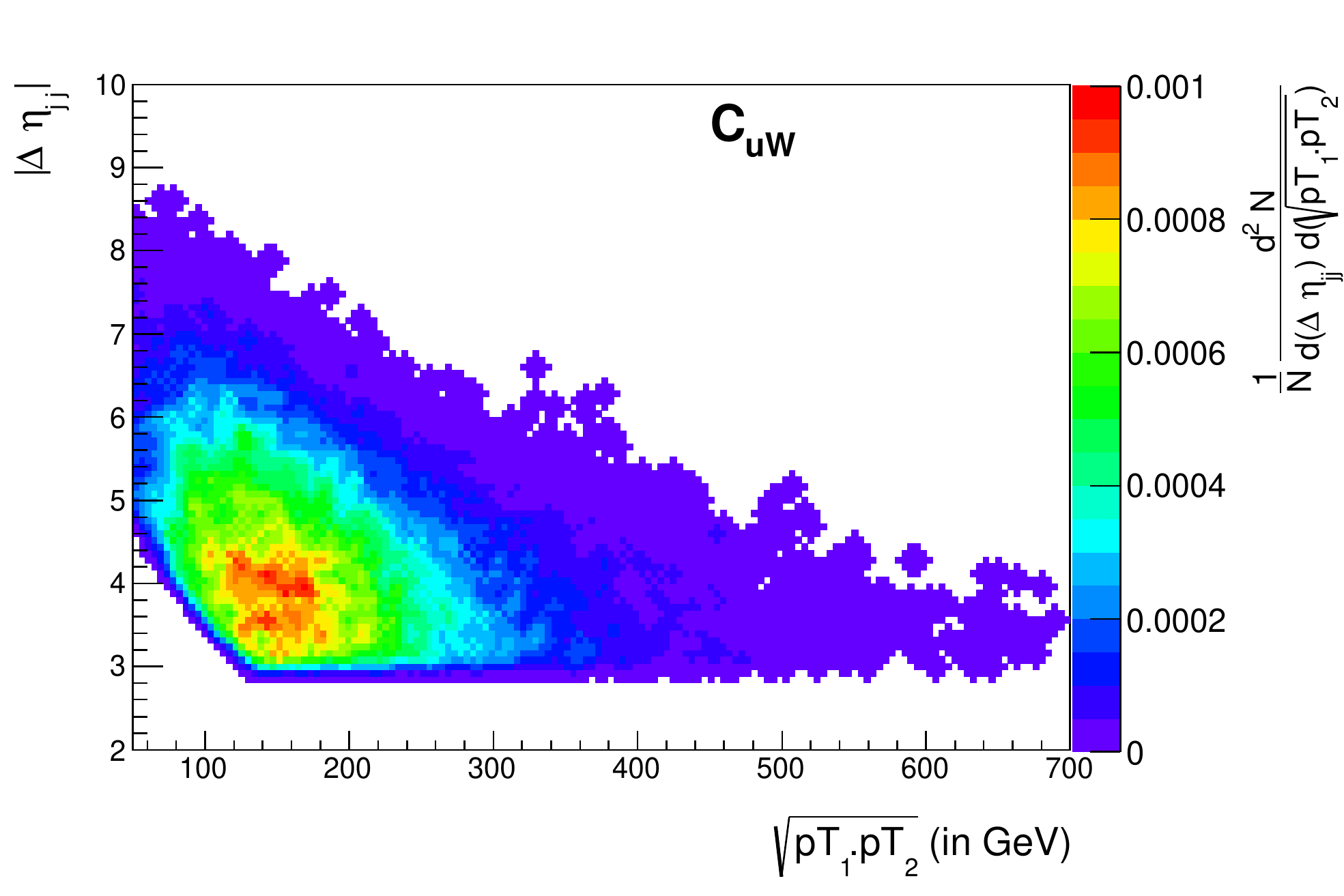} \\
			\hspace{4mm}(c)&\hspace{4mm}(d)
	\end{tabular}}
	\caption{Two dimensional histograms showing the correlation of the
		rapidity gap of jets $\Delta \eta_{jj}$ with $p_{T12}$, the geometric mean of $p_T$ of two leading jets at $\sqrt{s}=14$ TeV. The $z$-axis indicates the normalised frequency of events, in arbitrary units. The SM contributions in the absence of dimension-6 interactions have been subtracted in (b), (c) and (d). ${C \over {\Lambda^2}} = 0.3 ~\rm TeV^{-2}$ has been assumed in (b), (c) and (d).}
	\label{correlation}
\end{figure}

Next, we look at  the distribution of {\em rapidity separation of the forward jet pair} (Fig. \ref{tab:vbfkinematicdistrbn} (b)). The shape of the distributions are similar, however, with different normalisations due to interfering or non-interfering nature of the corresponding dimension-6 operators. Nevertheless, a careful look at this plot reveals, that events with smaller $\Delta \eta_{jj}$  ($< 4$) are mostly  generated by these higher dimensional operators  which implies that the events originating from new physics, are characterised with two forward jets with smaller rapidity separation at least for the operators $\mathcal{O}_{H q}^{(1)}$ and $\mathcal{O}_{uW}$. This  observation, in association with the fact that new physics events appear with high $p_T$ jets, can help us in separating the new physics rich phase space region from the SM. We demonstrate the correlation between two variables $\Delta \eta_{jj}$ and $p_{T12}$ in Fig. \ref{correlation}, where it is found that the populated regions in the $p_{T12}$ - $\Delta \eta _{jj}$ space display a shift when the new physics effects due to higher dimensional operators are included.

Before we delve into a discussion of such a correlation between $\Delta \eta _{jj}$ and $p_{T12}$, let us comment on the method that we have followed to obtain such a distribution. Although, new physics effects are visible in the $ p_{T12}$ distributions of the tagging forward jets, for high $p_{T12}$ values (see Fig. \ref{tab:vbfkinematicdistrbn}(a)), the difference between the SM and new physics is not prominent in the distribution of rapidity separation between the jets (see Fig. \ref{tab:vbfkinematicdistrbn}(b)). 
Due to a large cross-section of VBF Higgs production in the SM, any small modification due to the dimension-6 SMEFT operators, in the shape of kinematic  distributions, becomes less distinct, particularly in the $\Delta \eta_{jj}$ plot. It is these small relative differences between the EFT and the SM predictions that we are interested in, as they drive the sensitivity of new physics. Therefore, in order  to highlight any modification due to these new interactions, {\em we have subtracted, bin by bin, the number of events predicted purely by the SM}  from the total number of events obtained in presence of any of the aforementioned dimension-6 operators along with SM. We study the regions satisfying this criteria \cite{Baglio:2020oqu}. Experimentally, this amounts to subtracting the purely SM prediction from the experimental data, an exercise that is reasonably reliable in view of the extent studies on SM contribution to VBF. Effect of this subtraction is evident  as one can see the different position of peaks (red regions) of the two dimensional histograms. 

Some salient features of such a double differential  distribution are following:
\begin{itemize}
	\item  The region with moderate $p_{T12}$ ranging from 50 GeV to 400 GeV (with mean value at 100 GeV) and $\Delta \eta_{jj}$ around 5.2 (red colour indicates larger larger number of events), is mostly populated by SM VBF-Higgs events. The central region is depleted of any hadronic activity due to color singlet (in this case electroweak gauge bosons) exchange in t-channel.
	
	\item Regions with high values of  $ (> $ 100 GeV and extending upto 700 GeV)  and relatively smaller rapidity gaps ($\Delta \eta_{jj} \leq 4$) are populated by the new physics events. 
	
	\item It implies that these new vertices tend to push $p_T$ of jets to higher values and these hard jets are at small $\Delta \eta_{jj}$ gap compared to SM. Such a correlation is most prominent for $C_{uW}$ which have $|\Delta \eta_{jj}|$ distribution peaking around $4$. The effects of other two operators, too, extend to jet-$p_T$ values as high as 700 GeV. We urge the experimentalists to revisit the VBF data and ascertain or rule out the presence of events in the above region in the $p_{T12}$-$\Delta \eta_{jj}$ plane. 
\end{itemize}

\section{Sensitivity of  the VBF signal to  new interactions}
\label{sec4}
 We are now ready to quantify the sensitivity of VBF Higgs signal to the Wilson coefficients of the higher dimensional operators that we have been using 
  in our discussion. We calculated the projected significance in the vector boson fusion channel for illustrative values of ${C_i} \over \Lambda ^2$, for 14 TeV LHC at $3000$ fb$^{-1}$ luminosity. The significance $\cal Z$ \cite{cowan} is defined as follows:
		
\begin{equation}
{\cal Z} = \sqrt{2 [(S+B) \text{Log}\left(1+\frac{S}{B}\right) - S]}
\label{significance}
\end{equation}
 
 Signal $(S)$ is defined as $S = \vert N^{H}_{BSM} - N^{H}_{SM} \vert $.   Here, $N$ is the number of events for a given time integrated luminosity. We emphasise that generation of our signal events comprises of all the topologies (driven by the SM and EFT couplings) leading to the  $pp \to h(\to \gamma \gamma) jj$ final states.  Such final state topologies can arise from Higgs production via VBF,  in association with a $W/Z$ or via gluon fusion  process. Although these processes may interfere, the invariant mass of the two leading jets is itself a  powerful discriminating variable that permits us to exclusively select final states arising from VBF  mechanism.  The quantity $B $ is defined as $N^{H}_{SM} + N^{NH}_{SM}$.  $N^{H}_{BSM(SM)}$ in our signal consists of the number of VBF Higgs events in the SM. We have also included, in the VBF-enriched phase space, the number of SM Higgs events produced via gluon fusion channel and Vh channel that are allowed by VBF selection cuts and finally, $N^{NH}_{SM}$ is the number of non-Higgs events (leading to the same final state with two photons and two jets) in the SM \footnote{The non-Higgs backgrounds consist of non-resonant production of  di-photon,  single-photon and fake photons in association with more than one jet.} allowed by the VBF selection cuts.
      
  Higgs boson production via gluon fusion (ggF) and  in association with a W/Z  also contaminates  the VBF-Higgs cross-section. Rate of production of a Higgs boson via ggF and  passing through VBF selection criteria, is estimated to be 30\% \cite{ggFandvh} of  true VBF Higgs cross-section. In order to optimise our event rates, we impose a cut on rapidity gap between the tagged forward jets of 3 instead of 4  and an invariant mass of at least $600$ GeV for the tagged forward jet pair, instead of $400$  GeV used in \cite{ggFandvh}. We assume Higgs production cross-section due to ggF  passing through VBF selection criteria is 40\% that of true VBF cross-section. Higgs boson production in association with a $Z$ or $W$-bosons, can also contribute to VBF signal. However, a demand of high invariant mass  ($M_{jj} > 600$ GeV) of a pair of jets appearing in opposite hemisphere $(\eta_1 \eta_2 < 0)$  controls this background. Finally, we add another 40\% of true VBF cross-section to background $(B$) to also take into account for SM contribution from non-Higgs events producing photons and jets passing VBF selection cuts.  The number of events, one thus, arrives at 815 events at $3000 $ fb$^{-1}$ is larger than the expected background estimates of 780 events at $3000$  fb$^{-1}$ in \cite{Araz:2020zyh}. Therefore, if we have made an error,  it is on the conservative side.

   To see how  sensitive  $\cal Z$ is  to the Wilson coefficients, we estimate $\cal Z$, in two different ways. The first one of them is by plugging into Eqn. \ref{significance}, {\em the total cross-sections} of signal and background subjected to the cuts. In addition we calculate  $\cal Z$, by comparing signal strength with background in bins 
   of $p_{T12}$. In the following, both the results along with their implication will be presented.

  In Fig. \ref{signific_total} we present  the variation of $\cal Z$ with $\frac{C_i}{\Lambda^2}$ for an integrated luminosity of 3000 $fb^{-1}$. The key features emerging from Fig. \ref{signific_total} are as follows:

   \begin{figure}[h!]
   	\centering
  	\includegraphics[width=12.cm,height=10.cm]{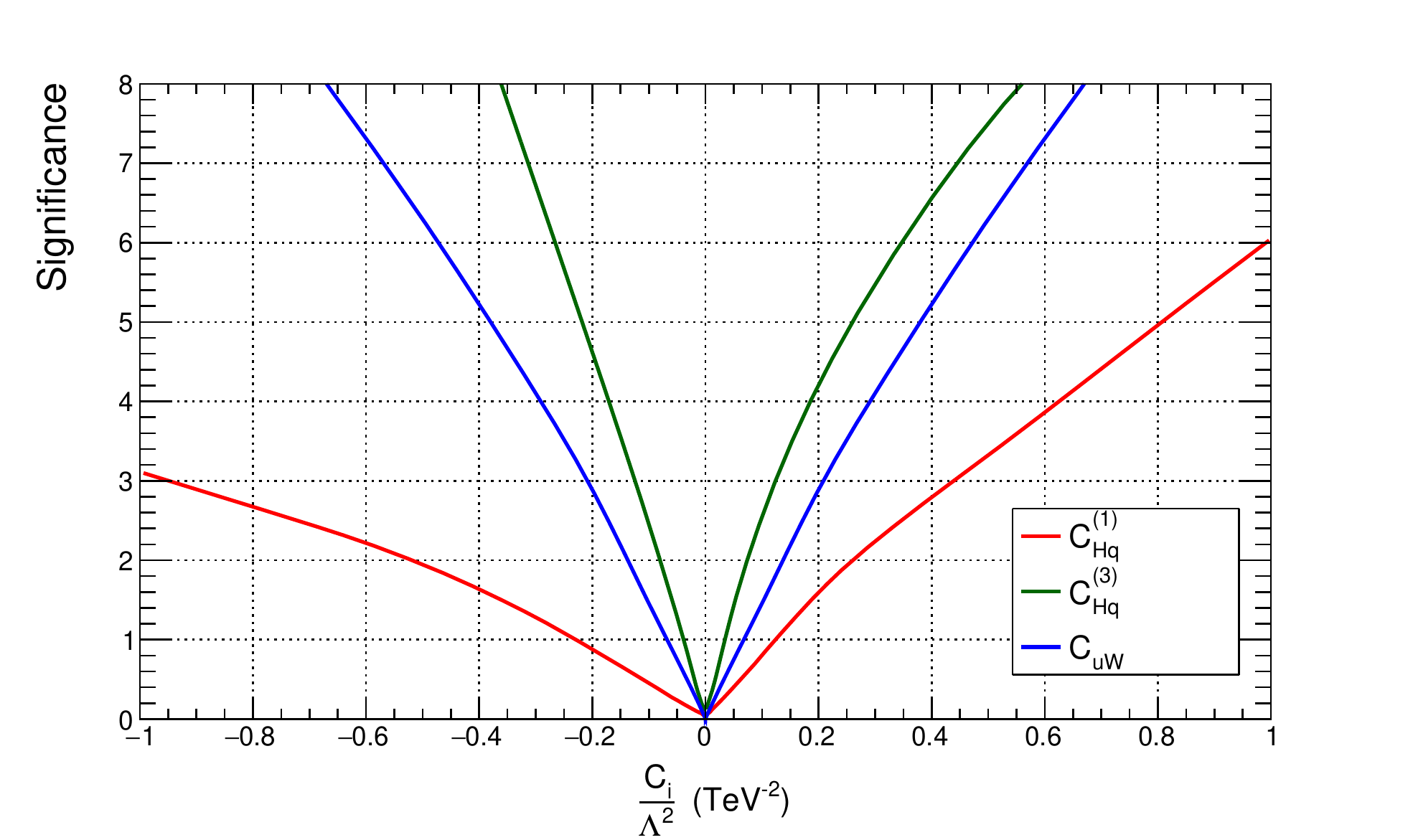}
   	\caption {Significance, $\mathcal{Z}$ (calculated using total cross-sections) as function of $\frac{C_i}{\Lambda^{2}}$  at $\sqrt{s} = 14~$TeV,  $3000 ~$fb$^{-1}$ for $p p \to h (\to \gamma \gamma)j j $}
	\label{signific_total}
   \end{figure}
   
      \begin{itemize}
   	\item Of all operators that we have considered, $\mathcal{O}_{H q}^{(3)}$ can modify the SM cross-section the most. Negative values of the Wilson coefficients of   $\mathcal{O}_{H q}^{(3)}$ increase the cross-section from its SM prediction more than the positive values. A $3\sigma$ signal significance over the background can be achieved even for small values of $ \frac{C_{Hq}^{(3)}}{\Lambda ^2} \in  (-0.13,0.14) {~\rm TeV}^{-2}$.
   	    	
   	\item The operator $\mathcal{O}_{H q}^{(1)}$  involves left-handed charged current of fermions with a weak gauge boson and Higgs. The positive values of Wilson coefficients of this operator enhances the cross-section more than negative coupling strength. Signal stands over the background a $3  \sigma$ level for $\frac{C_i}{\Lambda^{2}} \simeq 0.45~ \rm TeV ^{-2}$. Whereas, for $\frac{C_i}{\Lambda^{2}} = - 0.8 ~\rm TeV^{-2}$, 2.6$\sigma$ effect can be barely achieved.
   	
   	\item The dipole operator $\mathcal{O}_{uW}$ along with the SM results into the  largest cross-section of all three operators that we have considered. Despite having the lowest selection efficiency,  the cross-section is large enough for signal to stand against the background with  3$\sigma$ significance for $\vert \frac{C_i}{\Lambda^{2}} \vert > 0.22 ~{\rm TeV} ^{-2}$.	
	
	\item  Fig. \ref{signific_total} reveals $\vert \frac{C_{Hq}^{(3)}}{\Lambda^{2}} \vert < 0.14 {~\rm TeV}^{-2}$, which 
	seems to be a clear improvement over the limit derived on the same coupling from LEP EW data and Higgs data.  Similar but lesser  improvement of limits has been observed in case of $\frac{C_{Hq}^{(1)}}{\Lambda^{2}}$. $\frac{C_{uW}}{\Lambda^{2}}$ could not be constrained from LEP bounds. 
	
	  \end{itemize}

   Now we turn to the calculation of $\cal Z$ in the bins of $p_{T12}$. 
   Looking at the $p_{T12}$ distributions (in Fig. \ref{tab:vbfkinematicdistrbn}(a)), one can see that  the signal stands above the background  in individual bins (of $p_{T12}$) spanning over a wide range of its value. One can calculate the significance of signal in individual bins to gather maximum information from the  kinematics of the forward jets, we present in Table \ref{tab:vbfkinematicsig}, how signal significance changes along  the   bins of  $p_{T12}$ with different dimension-6 operators with the values of their Wilson coefficients set equal to $0.3$ TeV$^{-2}$ with an integrated luminosity of 3000 fb$^{-1}$. 
 The values of $\cal Z$, in same bins of $p_{T12}$,  are also presented in the same table, with ${C \over \Lambda^2 } = -0.3 ~\rm TeV ^{-2}$.  Such values of $\cal Z$  will help us to understand the effect of interference of new physics with the SM.
 
   \begin{table}[!hptb]
   	\begin{center}
   		\begin{footnotesize}
   			\begin{tabular}{| c | c | c | c |}
   				\hline
   				\hline 
   				& & &  \\
   				$~\rm bin$(GeV) & $\frac{C_{Hq}^{(1)}}{\Lambda^{2}} = 0.3 (-0.3) $ TeV$^{-2}$ &$\frac{C_{Hq}^{(3)}}{\Lambda^{2}} = 0.3 (-0.3)$ TeV$^{-2}$  & $\frac{C_{uW}}{\Lambda^{2}} = 0.3 (-0.3) $ TeV$^{-2}$    \\
   				& & &  \\
   				\hline
   				& & &  \\
   				150-200 & 3.57  (2.01) & 5.35 (9.03) & 2.61  \\
   				& & &  \\
   				\hline
   				& & &  \\
   				200-250 & 2.43 (1.67) & 4.07  (7.0) & 3.32 \\
   				& & & \\
   				\hline
   				& & & \\
   				250-300 & 2.28  (1.32)& 3.68 (4.7)& 4.19 \\
   				& & &  \\
   				\hline
   				& & &  \\
   				300-350 & 1.65 (1.09)& 2.45 (3.84) & 4.45 \\
   				& & &  \\
   				\hline
   				& & &  \\
   				350-400 & 1.54 (0.92) & 2.57 (3.35)& 5.53 \\
   				& & &  \\
   				\hline
   				& & &  \\
   				400-450 & 1.39 (0.78) & 2.15  (3.18)& 5.95  \\
   				& & &  \\
   				\hline
   				& & &  \\
   				450-500 & 1.24 (0.49) & 1.94 (2.65) & 6.32  \\
   				& & &  \\
   				\hline
   			\end{tabular}
   		\end{footnotesize}
   		\caption{Variation of signal significance, $\cal Z$ (calculated using Eqn. \ref{significance}) along the bins of $p_{T12}$ of tagging forward  jet pair  with {\em positive} and {\em negative} values of Wilson coefficients. }
   		\label{tab:vbfkinematicsig}
   	\end{center}
   \end{table}

The following points emerge from Table \ref{tab:vbfkinematicsig}:  
\begin{itemize}
\item For $\mathcal{O}_{H q}^{(1)}$ and  $\mathcal{O}_{H q}^{(3)}$, non-symmetric cut-efficiencies as function of $\frac{C}{\Lambda ^2}$ (see Fig. \ref{efficiency}) leads to the significance, $\cal Z$, to have different sensitivity to positive and negative values of $\frac{C}{\Lambda ^2}$. For example,  the operator ${\cal O}_{Hq}^{(3)}$ which interferes destructively with the SM, signal significance improves appreciably while calculated with $C_{Hq}^{(3)}  / TeV^{-2} < 0$ in comparison to its values calculated with $C_{Hq}^{(3)} / TeV^{-2}> 0$. Whereas for a constructively interfering operator $\mathcal{O}_{H q}^{(1)}$, a higher signal significance can be achieved always with $ C_{Hq}^{(1)} / TeV^{-2} > 0$.  Signal cross-section driven by $\mathcal{O}_{uW}$, does not show such sensitivity to the sign of its Wilson coefficient, as it does not interfere with the SM.

\item $\cal Z$ decreases monotonically along bins of increasing $p_{T12}$ for  the operators,  $\mathcal{O}_{H q}^{(3)}$ and $\mathcal{O}_{H q}^{(1)}$, while for $\mathcal{O}_{uW}$, $\cal Z$ steadily increases with $p_{T12}$. Significance, in each bin, is the joint outcome of how the SM and BSM contributions have their own $p_{T12}$ dependence and what their interplay is. Any cross-section in hadronic collision is a convolution of partonic cross-section with parton distribution functions (PDF). PDFs decrease with increasing  $p_{T12}$ (higher collisional energy). The {\em signal} rate at the parton level  either remains nearly independent of $p_{T12}$ (for $\mathcal{O}_{H q}^{(1)}$ and  $\mathcal{O}_{H q}^{(3)}$) or  increases (for $\mathcal{O}_{uW}$)  at a higher rate than the decrement of PDFs. Thus, in the latter case, enhancement of the EFT contribution with higher energy ($p_{T12}$) always improves the significance in high $p_{T12}$ bins.
\end{itemize}  
  \begin{figure}[h!]
   	\centering
   	\subfloat{	
   		\begin{tabular}{cc}
   			\includegraphics[width=8.1cm,height=7.1cm]{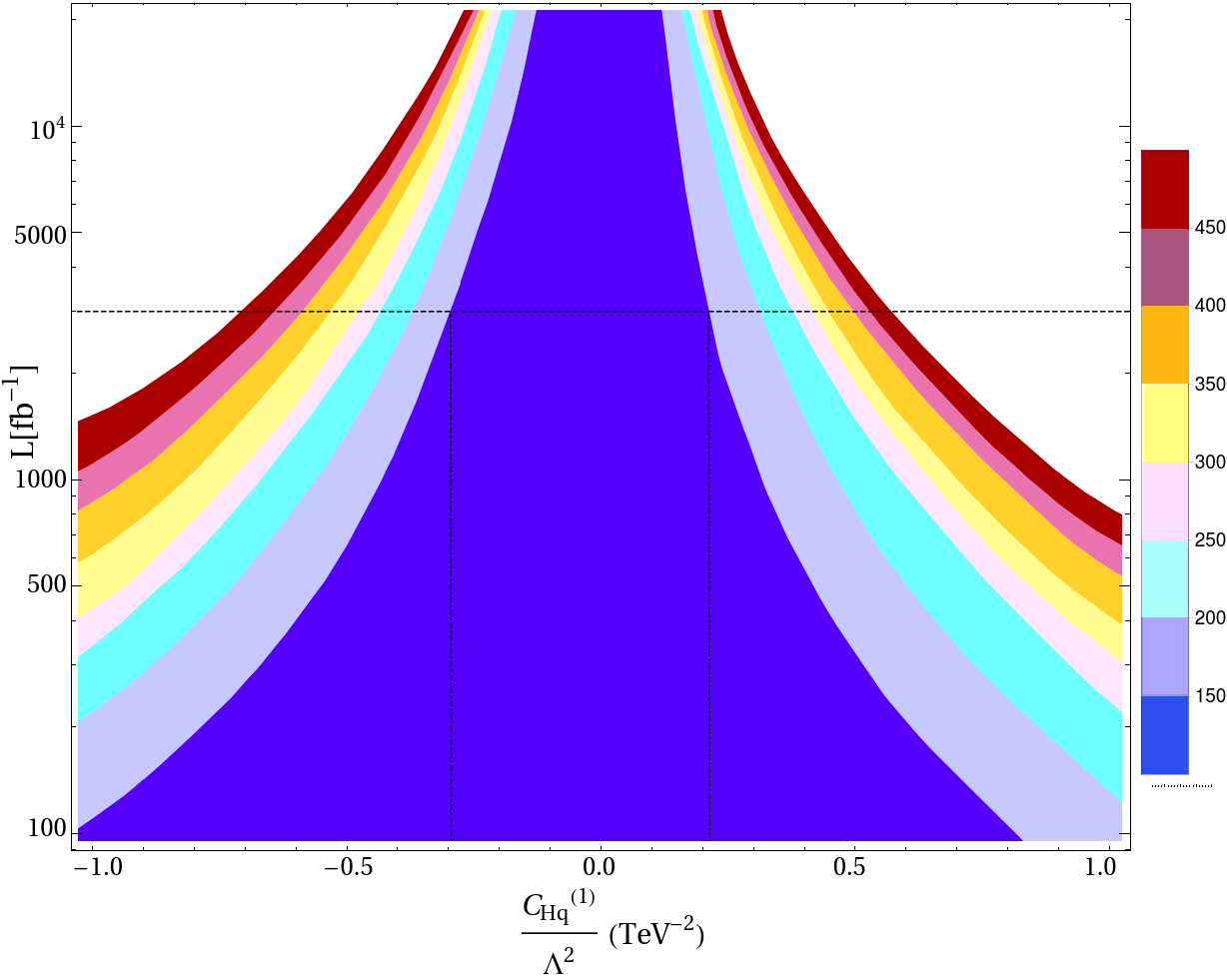} &
   		    \includegraphics[width=8.1cm,height=7.1cm]{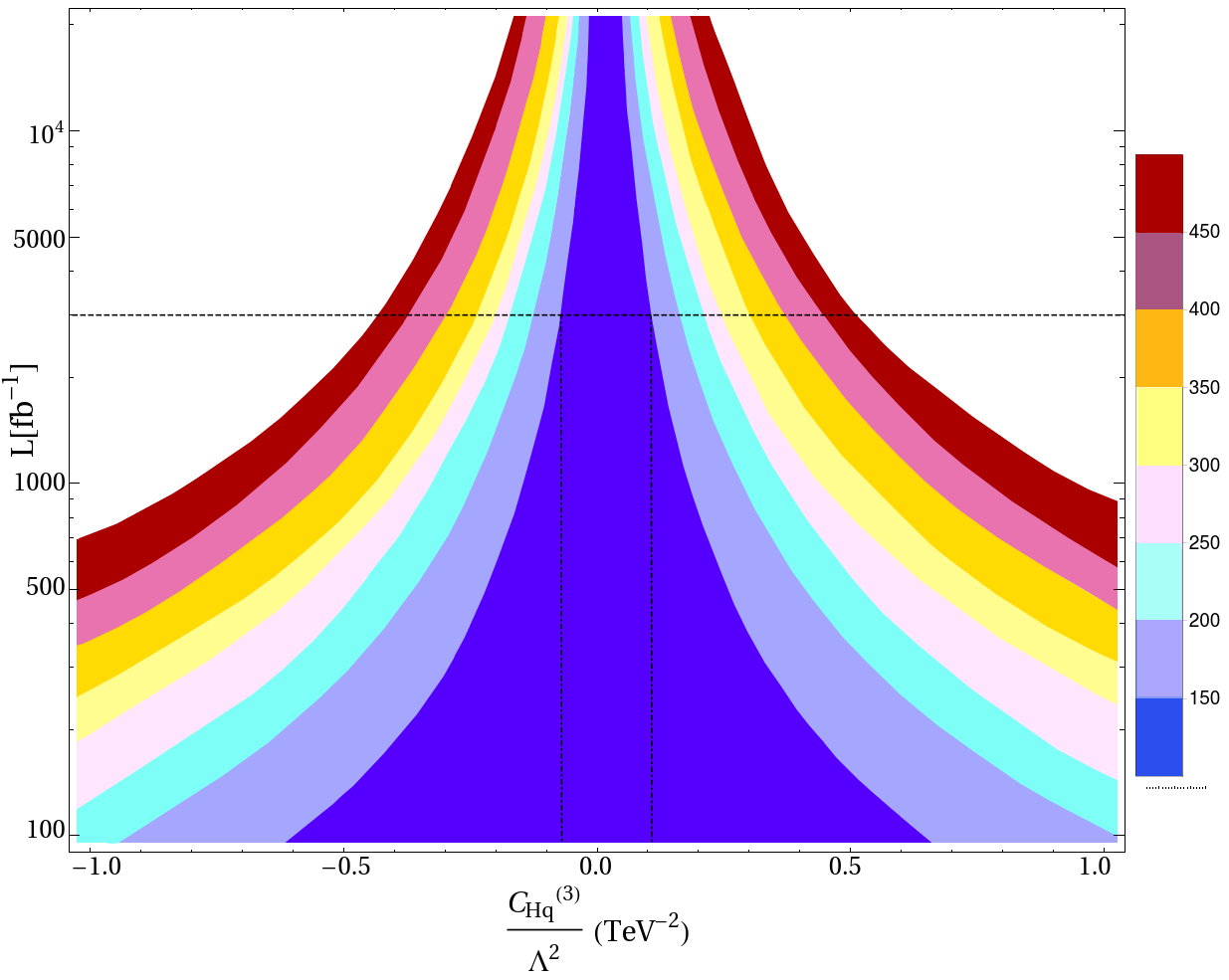}\\
   			\hspace{4mm}(a) &\hspace{8mm}(b) 
   	\end{tabular}}\\
   	
   	\begin{center}
   		\includegraphics[width=8.1cm,height=7.1cm]{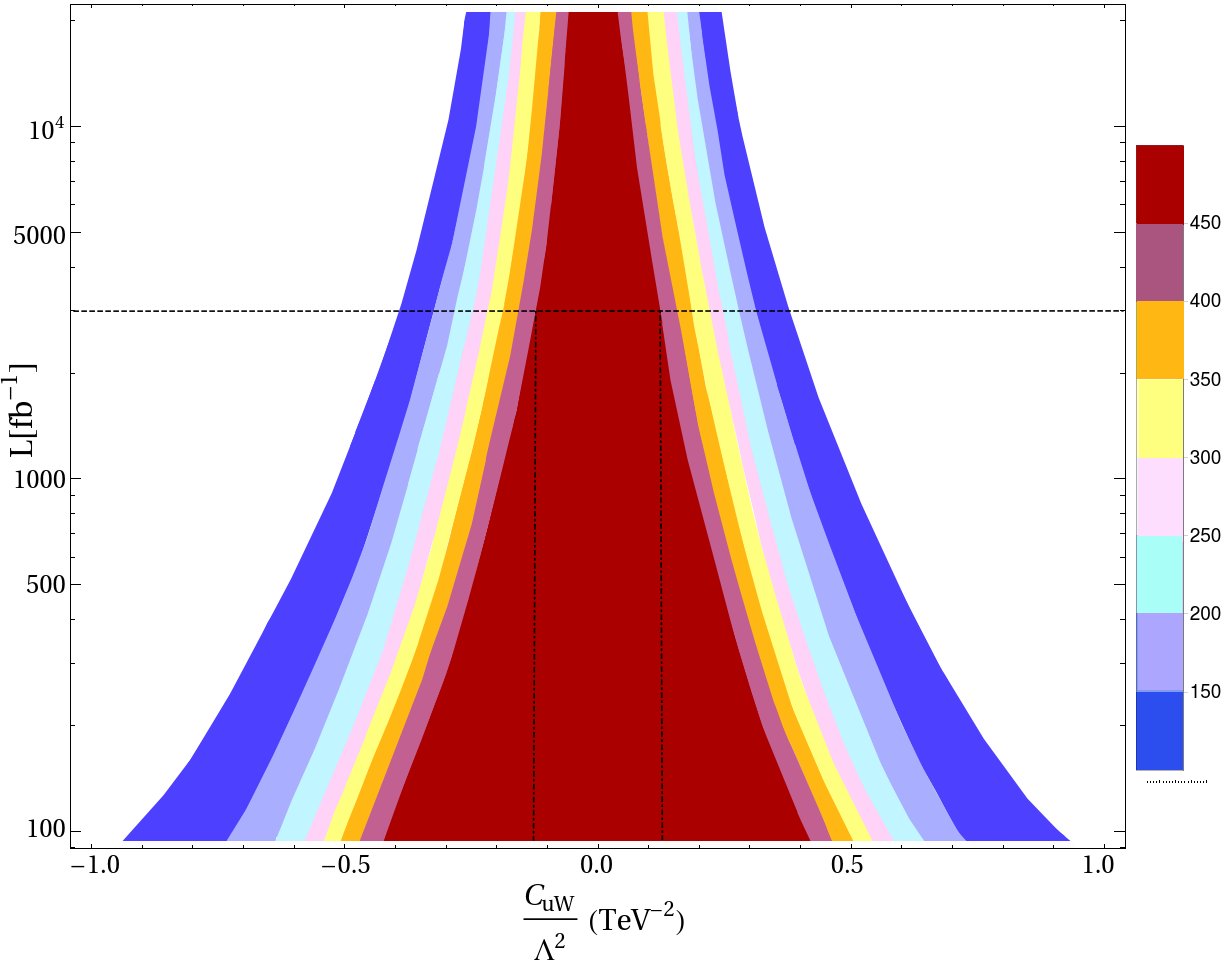} \\
   		\hspace{4mm}(c)
   	\end{center}
   	\caption{Required integrated luminosity at 14 TeV for achieving a 3$\sigma$ significance in bins of $p_{T12}$ as a function of Wilson coefficients of operators (a)$\frac{C_{Hq}^{(1)}}{\Lambda^2}$, (b) $\frac{C_{Hq}^{(3)}}{\Lambda^2}$ and  (c) $\frac{C_{uW}}{\Lambda^2}$. The different colored bands signify bins of $p_{T12}$ each of width 50 GeV.}
   	\label{reqdlumi}
   \end{figure}

 So far we have presented the signal significance  for a fixed integrated luminosity of 3000 $fb^{-1}$. However, we would also like to explore the luminosity required to obtain 3$\sigma$ exclusion limits on $C_i \over \Lambda^2$. To estimate the required
 luminosity, we have once again used the number of signal events in bins of $p_{T12}$, each of width 50 GeV, covering a range of 100 - 500 GeV for achieving $3 \sigma$ significance as a function of the Wilson coefficient. This is shown for three operators in Fig. \ref{reqdlumi}(a),(b),(c).  The vertical dotted dashed lines on each panel, represent the intervals of $C_i \over \Lambda^2$, which can be explored or ruled out at $3 \sigma$, with an integrated luminosity of 3000 $fb ^{-1}$, marked by a horizontal line on each panel.
 
 One can directly read from Fig. \ref{reqdlumi}, minimum luminosity required for signal with a given value of Wilson coefficient, to be greater than 
 3$\sigma$ fluctuation of background. Let us recall the 3$\sigma$ limits on $C\over \Lambda^2$ obtained by comparing total cross-section of signal to background.  $C_{H q}^{(1)} \over {\Lambda ^2}$ has been constrained between $-0.90 ~\rm TeV^{-2}$ and $0.45 ~\rm TeV^{-2}$ (see Fig. \ref{signific_total}) with 3000 fb$^{-1}$ of data. However,  Fig. \ref{reqdlumi}(a) tells us that a calculation of significance (with same luminosity) in the $p_{T12}$ bin of 100 - 150 GeV, could impose a more severe limit of $(-0.29 : 0.21) ~\rm TeV^{-2}$ on the same coupling. Similarly, the allowed region for  $C_{H q}^{(3)} \over {\Lambda ^2}$ becomes $(-0.09 : 0.15) ~\rm TeV^{-2}$, from the signal significance in the $p_{T12}$ bin of 100 - 150 GeV.
 Finally, the allowed region for $C_{uW} \over {\Lambda ^2}$ becomes $(-0.12 : 0.12) ~\rm TeV^{-2}$ calculated in the  $p_{T12}$ bin of 450 - 500 GeV.  A comparison of the two above methods, thus, emphasizes the usefulness of sensitivity information in individual bins of $p_{T12}$. This guides us to the most profitable bins in looking for effects of SMEFT.
        
  \begin{figure}[h!]
   	\centering
  	\includegraphics[width=6.cm,height=6.cm]{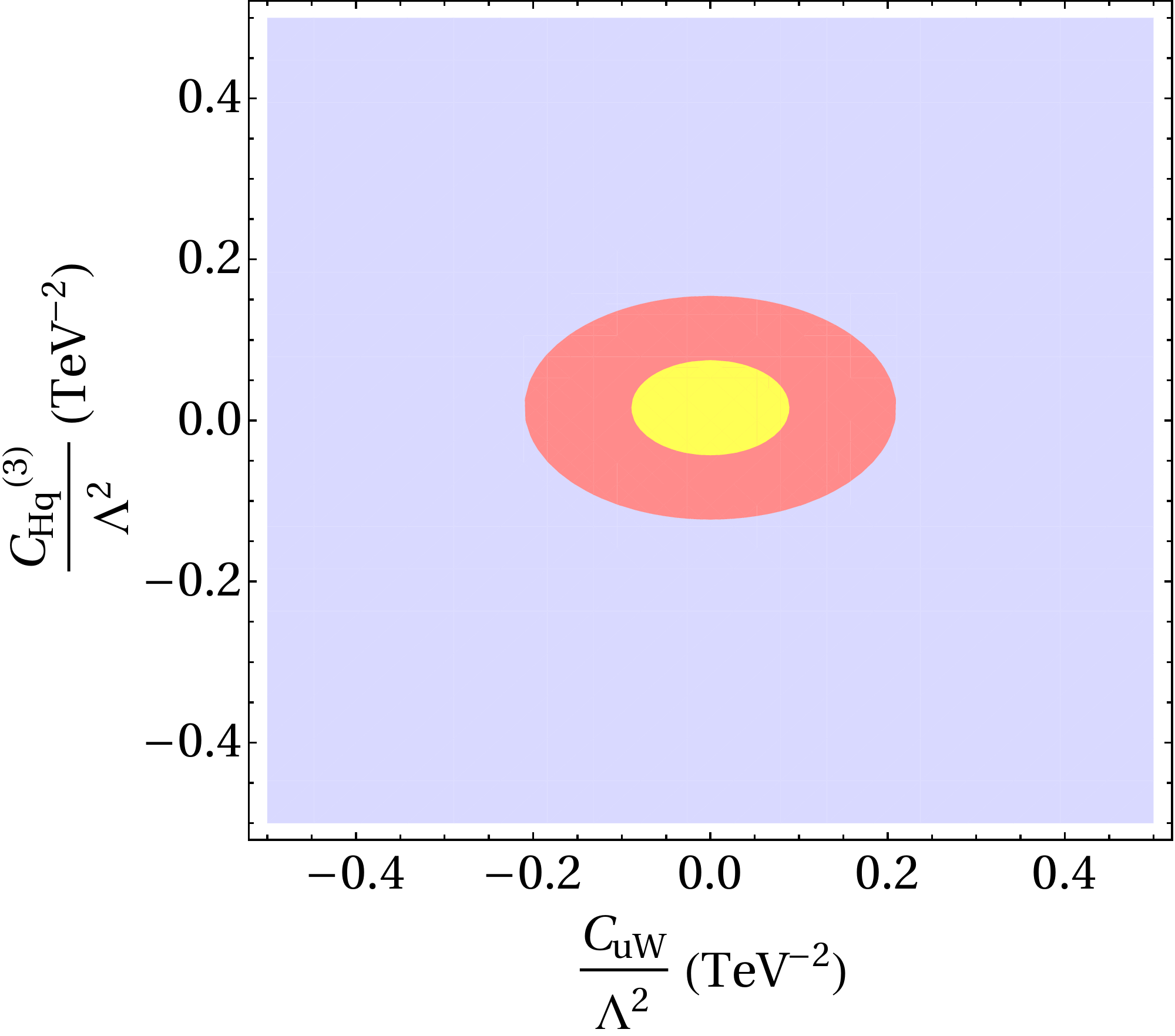}~~~~~
  	\includegraphics[width=6.cm,height=6.cm]{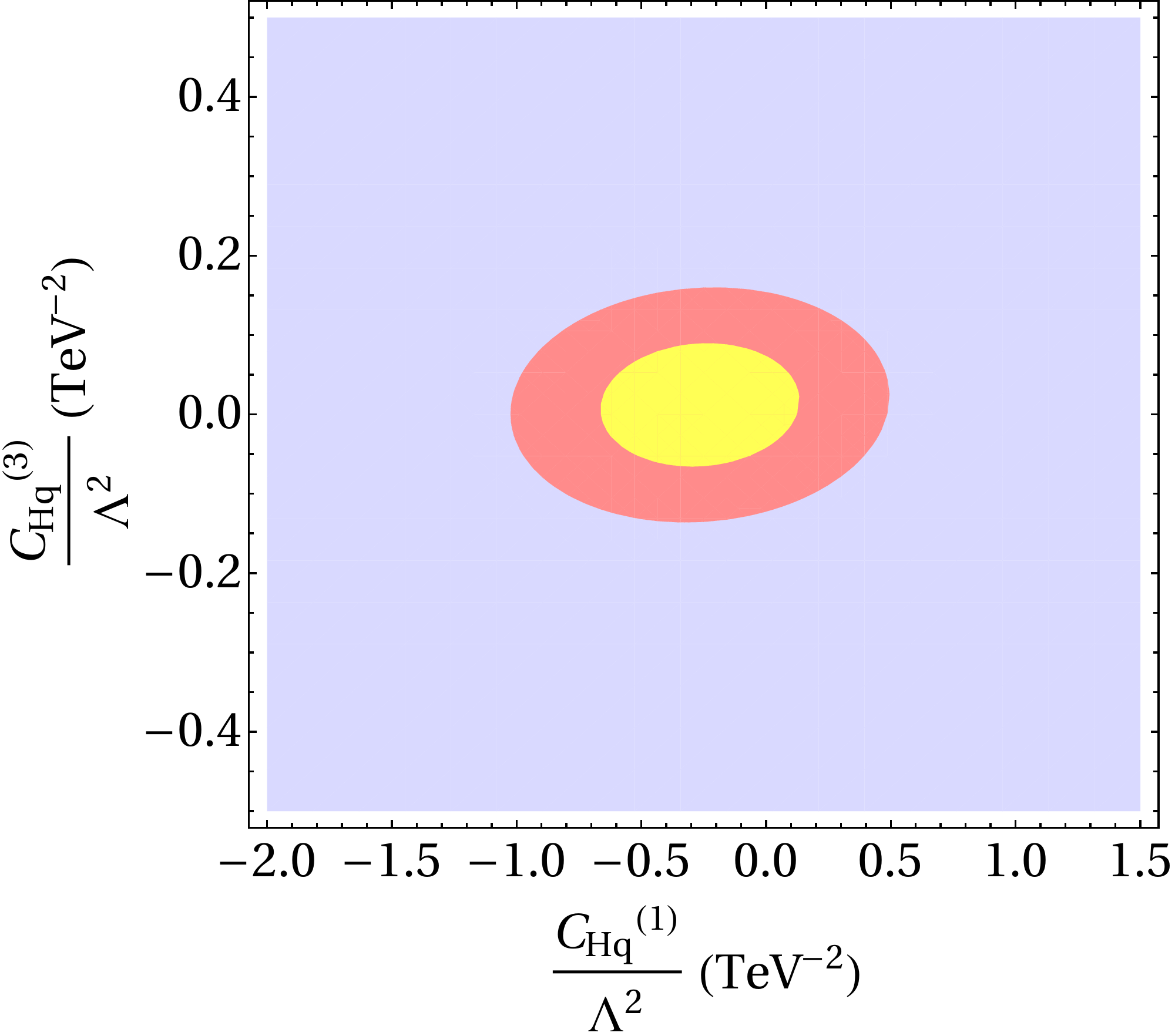}
   	\caption {Projections for a 3$\sigma$ signal significance in  blue shaded regions in the parameter space in (a) $\frac{C_{uW}}{\Lambda^2}$ - $\frac{C_{Hq}^{(3)}}{\Lambda^2}$ and (b) $\frac{C_{Hq}^{(1)}}{\Lambda^2}$ - $\frac{C_{Hq}^{(3)}}{\Lambda^2}$ planes with
	3000 $fb^{-1}$ data at 14 TeV run of the LHC. The 2$\sigma$ level approximating the 95\% confidence level exclusion bounds are shown in yellow shaded regions.}
	\label{contours}
   \end{figure}

We have presented our results assuming one non-zero dimension-6 operator at a time. Before closing this section,  let us discuss in brief,  the effect of two non-zero dimension-6 operators  on the expected sensitivity to the signal. In Fig. \ref{contours}, we have marked the regions in blue in $\frac{C_{uW}}{\Lambda^2}$ - $\frac{C_{Hq}^{(3)}}{\Lambda^2}$ plane, where  a signal significance of $3$ or more can be achieved. If no deviation is observed in the VBF Higgs production, we can obtain an upper limit on the Wilson coefficients at $2\sigma$ confidence level which yields bounds on the Wilson coefficients at the 95\% CL as shown in the yellow shaded region. A similar plot has been presented in the same figure with non-zero values of $\frac{C_{Hq}^{(1)}}{\Lambda^2}$ and $\frac{C_{Hq}^{(3)}}{\Lambda^2}$.    For both the plots, any possible pair of values of the relevant couplings chosen from the blue region, result into a signal away from 0 by at least 3 standard deviations.  The correlation between the couplings  shown in the plots can be understood by looking at the expression for the signal cross-section in the case when $C_{uW}$ and $C_{Hq}^{(3)}$ are taken non-zero at the same time:
\begin{equation}
\sigma(C_{uW}, C_{Hq}^{(3)}) = 0.653 + 0.27 \;(\frac{C_{uW}}{\Lambda^2})^2 + 0.167\;(\frac{C_{Hq}^{(3)}}{\Lambda^2})^2 - 0.391 (\frac{C_{Hq}^{(3)}}{\Lambda^2})
\label{eq_corr}
\end{equation}

For negative values of $C_{Hq}^{(3)}$, interference term (linear in $C_{Hq}^{(3)}$) adds to the quadratic $(C_{Hq}^{(3)}) ^2$ term and 
a relatively smaller $C_{uW}$ can achieve $3\sigma$ signal significance. On the other hand, for $C_{Hq}^{(3)}/\rm TeV^{2} > 0$, this destructive interference  between the SM and $O_{Hq}^{(3)}$ will be compensated by both $(C_{uW})^{2}$ and $(C_{Hq}^{(3)})^{2}$ terms, meaning that the $\frac{1}{\Lambda^4}$ contribution helping to achieve  a 3$\sigma$ (or more) signal significance. The interval $|{C_{uW} 
\over {\Lambda^2} }| > 0.22~\rm TeV^{-2}$ and $| {C_{Hq}^{(3)} \over {\Lambda^2}}| > 0.15~\rm TeV^{-2}$ corresponds to a cross section of VBF Higgs production in the di-photon channel that is 7\% away from the SM.

Similarly, in  ${C_{Hq}^{(1)} \over {\Lambda^2}}-{C_{Hq}^{(3)} \over {\Lambda^2}}$ plane, a correlation exists between the two. For $C_{Hq}^{(1)}/\rm TeV^{2} < 0$, the nature of its destructive interference with SM, will be outweighed by the strong interference of $C_{Hq}^{(3)}$ in this range and until ${C_{Hq}^{(1)}  \over {\Lambda ^2}}< -1~\rm TeV^{-2}$, its quadratic dependence takes over and adds to the total rate of the process. For $C_{Hq}^{(1)}/\rm TeV^{2} > 0$, although it interacts constructively with SM, the VBF process is more sensitive to $C_{Hq}^{(3)}$ and its destructive interference asks for a greater value of $c_{Hq}^{(1)}$. The above mentioned pair of coefficients allow a region of parameter space which has, for instance, at ${c_{Hq}^{(1)} \over {\Lambda ^2}} = 0.4~\rm TeV^{-2}, {C_{Hq}^{(3)} \over {\Lambda ^2}} = -0.2~\rm TeV^{-2}$, a cross-section 6\% away from the SM value.

One can also check the effect of two  operator couplings varying at the same time while estimating the signal significance in individual bins of $p_{T12}$. For instance, with 3000 $fb ^{-1}$ data, a 3$\sigma$ difference can be achieved with ${C_{uW} \over {\Lambda ^2}}= \pm \;0.23  ~\rm TeV^{-2}$ (while keeping the values of other couplings to zero) in the $p_{T12}$ bin of $300 -350$ GeV (see Fig. \ref{reqdlumi}(c)).  The effect of turning on $C_{Hq}^{(3)}$ along with $C_{uW}$ can be easily understood from Eqn. \ref{eq_corr}. With negative  $C_{Hq}^{(3)}$,  a 3$\sigma$ effect can easily be achieved in the same bin with a smaller value of $C_{uW} \over {\Lambda ^2}$ than $0.23 ~\rm TeV^{-2}$. 

All the results involving the dimension-6 operators, presented above, have been derived on the basis of LO estimation of cross-section. At this end, we would like to comment on the possible inclusion of NLO QCD corrections to the new physics cross-sections. We have estimated the NLO QCD corrected cross-section for VBF process involving the operators ${\mathcal O}_{Hq}^{(1)}$ and ${\mathcal O}_{Hq}^{(3)}$ which are available within the {\tt{SMEFT@NLO}} \cite{smeft_nlo} package.  Following inferences can be drawn from our analysis. Firstly, the total cross-section at the NLO, as compared to LO, always goes up by the order 12 - 25\%. The $k$-factors for VBF process, calculated (from the total cross-section) including  ${\mathcal O}_{Hq}^{(1)}$ and  ${\mathcal O}_{Hq}^{(3)}$  operators are $1.12$ and $1.25$ respectively assuming ${C_i \over \Lambda^2 } = 0.3~\rm TeV^{-2}$. Furthermore, the overall orientation and shapes of the distributions $(p_{T12})$ is not significantly altered. Thus, our LO estimates are conservative in nature. We have also checked that uncertainty in cross-section due to factorisation and renormalisation scale choice at NLO  is less than a percent when we change the scale from $m_h$ to $m_h \over 2$. The differential k-factors for both the operators are always greater than 1, becoming larger at high $p_{T12}$ bins, reaching $~1.25$ and $ 1.29$ in $p_{T12}$ bins of  400-450 GeV, for $C_{Hq}^{(1)}$ and $C_{Hq}^{(3)}$, respectively. In cases, where the interference with the SM is constructive (depending on the sign of the Wilson coefficient of the operator), the bin-by-bin distinguishability with the pure SM contributions improves in most bins when NLO effects are included.  In case of destructive interference, the distinguishability (from the SM) is adversely affected in the bins ranging from 150 - 350 GeV.  However, there are always several bins where a significance above $3\sigma$ have been achieved with NLO cross-section. In general, high $p_T$ regions give better distinguishability on the inclusion of NLO effects.
      
\begin{figure}[h!]
	\centering
	\includegraphics[width=12.5cm,height=10.cm]{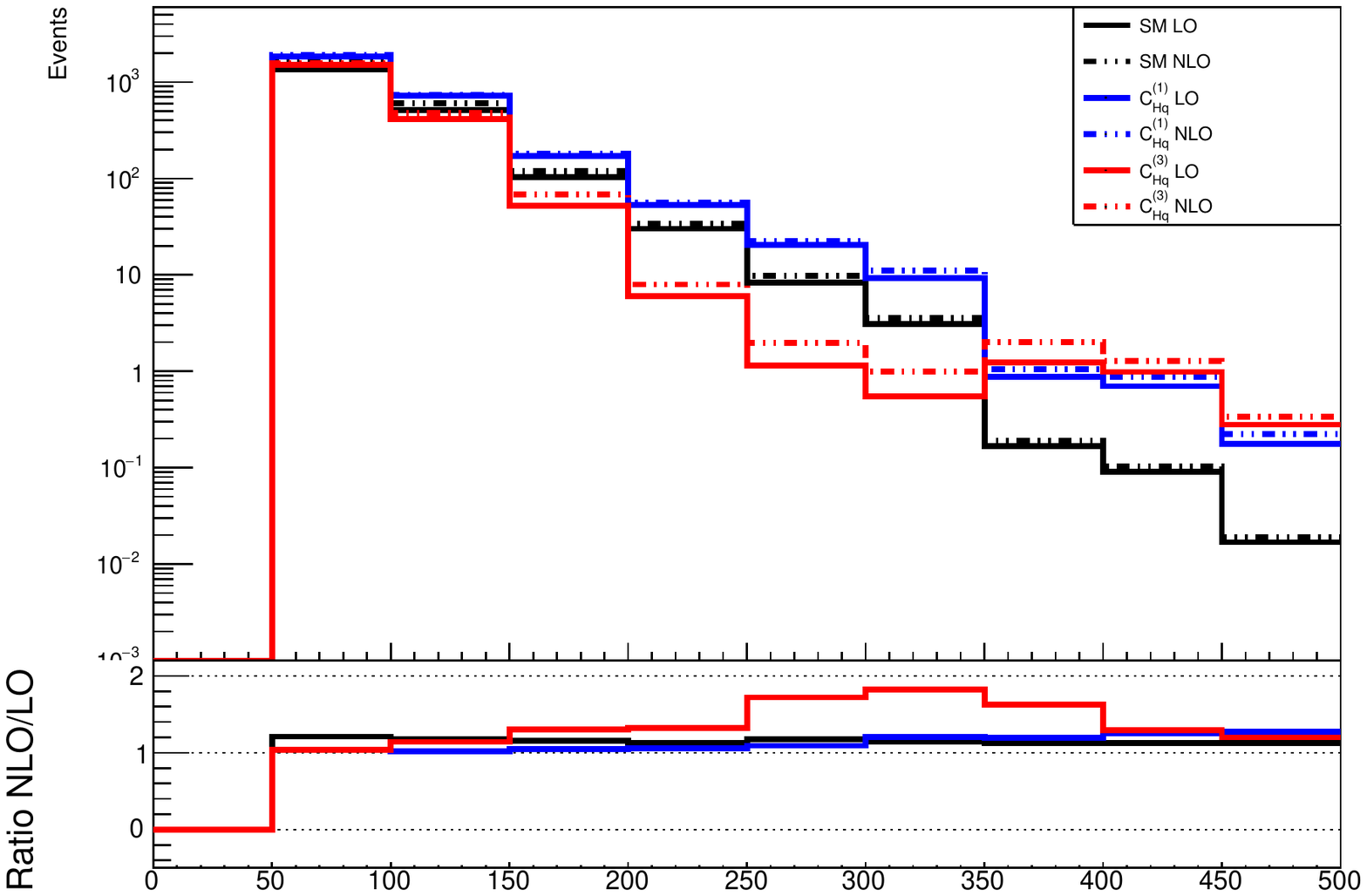}
	\caption {Distribution at LO (solid) and NLO (dotted-dashed) new interactions with $C_{Hq}^{(1)}$ (blue) and $C_{Hq}^{(3)}$ (red) benchmark points ($= 0.3$ TeV$^{-2}$) , in bins of $p_{T12}$, for $pp \to h(\to \gamma \gamma)jj$ at an integrated luminosity of 3000 fb$^{-1}$, 14 TeV LHC.}
	\label{nlo_geomeanpt_dist}
\end{figure}  
\section{Summary and Conclusion}
\label{sec5}
We have investigated the effects of some illustrative dimension-6 operators involving interaction of two quarks, Higgs field and a gauge boson on Higgs production via the vector boson fusion channel at the Large Hadron Collider.  To begin with, we have obtained the upper limits on the Wilson coefficients of the aforementioned operators by a simple unitarity analysis of the process $qq \rightarrow h Z, qq \rightarrow h W$. The values of the Wilson coefficients used in our analysis are consistent with all the erstwhile experimental results including weak universality, electroweak precision tests and LHC data. Parametrising the strength of the new interactions by the coefficients $C_i ~(C_{Hq}^{(1)}, C_{Hq}^{(3)},C_{uW})$, after a detailed cut-based Monte Carlo analysis, we study how the efficiencies of different acceptance cuts are altered for various values of $C_i$.
 
The final state considered in our study is di-photon in association with two forward jets. 
We have utilised the jet kinematics to distinguish the signal from background.   Our analysis does not depend on any particular decay mode of Higgs boson. To be more specific, we find that the presence of these dimension-6 operators would result into harder $p_T$ spectra of the two forward jets. Consequently, a harder $p_{T12}$ spectra will emerge on  inclusion of the effective operators discussed here. The present analysis has revealed a region in $p_{T12}$ - $\Delta \eta_{jj}$ phase space where  the cases in presence of dimension-6 operators along with the SM are populated with high energetic jets and less separated in rapidity direction. These regions define the new corners of phase space where SM is highly depleted. Particularly, the operator with explicit momentum dependence affects the rapidity between the two leading jets and enhances their transverse momenta, most prominently. The other two operators that interfere with the SM, also show similar effects to a lesser extent. 

The VBF Higgs signal can be mimicked by processes like Higgs boson production with two jets via gluon fusion,  and non-Higgs background like di-jet production with di-photon. With these backgrounds constituting nearly 80\% of  true SM VBF cross-section,  we have computed the significance  of our signal in two different ways: (i) by comparing the total cross-section of signal and background and (ii) by comparing the signal and background event rates in the bins of $p_{T12}$. Significant improvement  has been observed in obtaining  3$\sigma$ limits on $C \over \Lambda ^2$ when significance calculation has been done in separate bins of $p_{T12}$.  The projected 3$\sigma$ upper limits for integrated luminosity of 3000 fb$^{-1}$ from our analysis seem to be more restrictive than bounds coming from precision electroweak observables.

Several other kinematic observables can also be constructed out of the forward jets.  We have specifically checked that in distributions  of (i) {\em azimuthal angle difference of leading jet pairs}  (ii) {\em $\eta$-centrality} and (iii) {\em $p_T$ difference of leading jet pair} distribution, sufficient modifications to SM predictions can be observed with moderate values of Wilson coefficients of these higher dimensional operators.  A study including these additional kinematic variables and their possible correlation, by going beyond the standard cut-based approach will be reported in a follow-up study where other Higgs decay channels are also being taken into account.


{\bf Acknowledgements:} We thank Satyaki Bhattacharya for useful discussions. TB acknowledges the support from Council of Scientific and Industrial Research, Government of India. TB is thankful to Nabanita Ganguly and Tathagata Ghosh for insightful comments. Authors thank Satyanarayan Mukhopadhyay for an helpful discussion on the validity of EFT series.

 	
\end{document}